%% file: 00_farland.tex
\newcount\draft\draft=1 

\documentclass{article}
\usepackage{arxiv}
\usepackage{amsmath}

\usepackage{tikz}

\usepackage{float}
\input{01_customization}

\begin{document}

\date{}

\input{02_metadata}

\maketitle

\input{03_abstract}

\input{04_rights}

\input{05_introduction} 
\input{06_related_work}

\input{07_motivation}

\input{08_game_abstraction}
\input{09_architecture}
\input{10_developing_agents}

\input{11_evaluation}

\input{12_conclusion}


\bibliographystyle{plain}
\bibliography{00_farland}

\end{document}

%% file: 01_customization.tex
\usepackage{enumerate}
\usepackage{amssymb}

\usepackage{bm}
\usepackage{soul}
\usepackage[normalem]{ulem}

\newcommand{\E}{\mathbb{E}}

\ifcsname G\endcsname%
    
\else%
    
\fi%
\ifcsname C\endcsname%
    
\else%
    
\fi%

\renewcommand{\S}{\mathcal{S}}
\newcommand{\A}{\mathcal{A}}

\renewcommand{\P}{\mathcal{P}}

\ifnum\draft>0
  
\else 
  
\fi

\ifnum\draft>1
  \newcommand{\removedtext}[1]{\textcolor{red}{\sout{#1}}}

\else

  \newcommand{\removedtext}[1]{\ignorespaces}
\fi

\usepackage{tabu}
\usepackage{wrapfig}
\usepackage{graphicx}
\usepackage[all]{xy}

\usepackage{color}
\usepackage{fancyvrb}

\usepackage{cleveref}
\crefname{section}{§}{§§}
\Crefname{section}{§}{§§}

\newcommand\blfootnote[1]{%
  \begingroup
  \renewcommand\thefootnote{}\footnote{#1}%
  \addtocounter{footnote}{-1}%
  \endgroup
}


%% file: 02_metadata.tex
\title{\Large \bf Network Environment Design for Autonomous Cyberdefense}

\author{
{\rm Andres Molina-Markham, Cory Miniter, Becky Powell}\\
The MITRE Corporation
\and
{\rm Ahmad Ridley}\\
National Security Agency
} 

%% file: 03_abstract.tex
\begin{abstract}

Reinforcement learning (RL) has been demonstrated suitable to develop agents
that play complex games with human-level performance. However, it is not
understood how to effectively use RL to perform cybersecurity tasks.
To develop such understanding, it is necessary to develop RL agents using
simulation and emulation systems allowing researchers to model a broad class of
realistic threats and network conditions. Demonstrating that a specific RL
algorithm can be effective for defending a network under certain conditions may
not necessarily give insight about the performance of the  algorithm when the
threats, network conditions, and security goals change. 
This paper introduces a novel approach for network environment
design and a software framework to address the fundamental problem that network
defense cannot be defined as a single game with a simple set of fixed rules. We
show how our approach is necessary to facilitate the
development of RL network defenders that are robust against attacks aimed at the
agent’s learning. Our framework enables the development and simulation of
adversaries with sophisticated behavior that includes poisoning and evasion
attacks on RL network defenders.
\end{abstract}

%% file: 04_rights.tex
\blfootnote{Approved for Public Release; Distribution Unlimited. Public Release Case Number 21-0162. 
This technical data deliverable was developed using contract funds under Basic
Contract No. W56KGU-18-D-0004. The view, opinions, and/or findings contained in
this report are those of The MITRE Corporation and should not be construed as an
official Government position, policy, or decision, unless designated by other
documentation. \copyright 2021 The MITRE Corporation. ALL RIGHTS RESERVED.}

%% file: 05_introduction.tex
\section{Introduction}

Since the introduction of Deep Reinforcement
Learning~\cite{mnih_playing_2013}, numerous systems have successfully combined
fundamental ideas from Reinforcement Learning (RL) and Deep Learning to solve
hard problems with superhuman performance. However, it remains a challenge to
translate such progress into successfully using RL to tackle
complex tasks, such as those necessary for network defense. One fundamental
problem is that network defense cannot be defined as a single game with a simple
set of rules. Rather, proficient network defense corresponds to mastering a
spectrum of \emph{games} that depend on sets of (i) adversarial tactics,
techniques, and procedures (TTPs); (ii) quality of service goals and
characteristics of a network; and (iii) actions available for defenders and
their security goals.

In this paper, we present \emph{FARLAND}, a \emph{framework for
advanced Reinforcement Learning for autonomous network defense}, that uniquely
enables the \emph{design of network environments} to gradually increase the
complexity of models, providing a path for autonomous agents to increase their
performance from apprentice to superhuman level, in the task of reconfiguring
networks to mitigate cyberattacks. Our approach is aligned with research, such
as unsupervised environment design~\cite{dennis_emergent_2020} and automated
domain randomization~\cite{openai_solving_2019}, which highlight the need to
gradually scale the difficulty of a task to allow progress when learning to
master complex jobs. Importantly, FARLAND's abstractions for network
environment design include adversarial models targeting the RL-enabled decision
making component. This reflects the ultimate goals of FARLAND, which include
enabling the development of practical approaches for autonomous network defense;
and the evaluation of the robustness of such approaches against RL-targeted
attacks.

While a number of systems and frameworks already enable
research in RL---such as Gym~\cite{brockman_openai_2016}, DeepMind Lab
~\cite{beattie_deepmind_2016}, or Malmo~\cite{johnson_malmo_2016}---none are
adequate for performing research to apply RL for practical network defense. It
is not sufficient to build a system to simulate attacks, recognizing that it is
difficult to learn to reconfigure a real network by directly experiencing a
large volume of attacks in real environments. It is important to use a framework
to addresses the open problem of developing strategies to guide an agent through
a sequence of increasingly harder problems to guarantee progress (i.e. analog to
automated domain randomization~\cite{openai_solving_2019} or unsupervised
curriculum learning~\cite{dennis_emergent_2020}). Furthermore, such a framework
needs to expose an appropriate set of knobs (model parameters) during the
process of gradually increasing model complexity to tackle network defense.
FARLAND provides a solution to these challenges, and it is therefore unique in
enabling network environment design, an essential aspect for research toward
understanding how to apply RL for network defense.

Our work not only proposes concrete aspects that must be
 parametrized in the process of developing network environments, but also, it
 provides concrete approaches for controlling these parameters while learning to
 reconfigure a network for cyberdefense. Furthermore, we put forward a dual
 layer architecture to trade model fidelity for dramatically faster experience
 replay. Concretely, FARLAND allows the specification of (a) network models
 (which include gray (i.e., normal user) and red agent behavior) via generative
 programs~\cite{cusumano-towner_gen:_2019}, thus providing an answer to the
 problem of modeling environments with varying degrees of complexity and
 dynamics that are probabilistic and partly observable; and (b) sets of
 blue-agent actions, views of network state (observations), and reward
 functions, to increasingly enable more sophisticated blue policies. In
 addition, FARLAND allows researchers to develop compatible simulation and
 emulation network environments, such that games can be simulated (via
 generative programs) with a fraction of the computation cost required to
 emulate the corresponding game in a full software defined network, with real
 hosts, and control and data planes. Emulation network environments can be used
 to estimate the parameters of probabilistic models used in simulation; and to
 validate the performance of learned reconfiguration policies.

Other key considerations
 informed the design of FARLAND and contributed to its novelty. Concretely, FARLAND:

\textbf{1. Leverages state-of-the-art RL algorithm implementations and
distributed computing frameworks for AI:} Training an intelligent network
defender is computationally expensive. Simulating network-defender experiences
requires a large number of general-purpose computations. In addition, training
requires solving computationally intensive optimization problems through the
effective use of GPUs.

FARLAND allows researchers to scale the size and complexity of their networks
and models by leveraging RLLib~\cite{liang_ray_2017}, which abstracts common
distributed computing patterns that allow simple scaling when more resources are
available. 
FARLAND's architecture decouples tasks related to the design of RL algorithms
from tasks related to network environment design and other low
level aspects of how to coordinate the execution of multiple agent experiences,
the collection of data, etc.
While, RLLib includes implementations of about a
dozen RL algorithms~--e.g., A2C, A3C, DQN, DDPG, APEX-DQN, and IMPALA~--we argue
that there is a further need to develop novel algorithms that are robust against
attacks to manipulate AI-enabled decisions (i.e., evasion and poisoning
attacks). RLLib's architecture allows experts in algorithm design to cleanly
separate this task from problems related to distributed computing, as well as
from addressing environment modeling problems. 
FARLAND's abstractions also
separate the problems of defining  security goals, network and
adversarial models, from the problem of implementing a simulator or emulator to
effectively turn these models into an environment with which the learning agent
can interact.

\textbf{2. Supports cyber defense research:} Unlike existing systems that aim to
enable research in
RL~\cite{brockman_openai_2016,beattie_deepmind_2016,johnson_malmo_2016}, FARLAND
is specifically designed to support the development of (blue) RL agents that
learn to defend networks, performing actions that modify the configurations of
software-defined networks. 
While prior research aims to apply RL to solve networking
problems~\cite{chen_auto:_2018, ruffy_iroko:_2018}, FARLAND would be among the
first to provide an open framework specific for network
defense\footnote{CybORG~\cite{baillie_cyborg_2020} is also among one of the first.
However, our goals differ in significant ways~\cref{sec:related_work}.}. 

We argue that research should not focus on demonstrating that an approach can
enable the learning of policies for an agent under specific conditions because
network models, threat models, and security goals can vary dramatically.
Instead, researchers need the ability to model broad classes of models and
subject their approach to relevant variations of their models to understand the
implications of changing network conditions, assumptions about their
adversaries, security goals, and defender action spaces. 

Moreover, it is highly likely that real-world network defense is actually
implemented by multiple agents, each in charge of a simple set of decisions from
only a subset of all data that can be observed. Hence, context will influence
engineering efforts to determine suitable observation spaces, action spaces, and
reward functions. Our work aims to provide an experimentation framework to
investigate these issues rather than to provide a solution to the problem of
creating an autonomous network defender that is good in a small set of concrete
situations.

\textbf{3. Facilitates transition from research to practice:} With the goal of
closing the gap between research and practice, FARLAND’s coherent
simulation-emulation paradigm offers researchers an accelerated development of
RL agents via simulation, while allowing for emulation-supported refinement and
validation. The behavior of agents (red, gray, and blue) can be simple or
sophisticated. Researchers can start by modeling simple behaviors to enable
rapid progress and fundamental understanding. However, the modularity of FARLAND
provides a path for extending the behavior complexity of agents. Our goal to
close the gap between theory and practice differs from the goals of other
systems to provide environments that offer a broad range in complexity, partial
observability, and multi-agency (e.g., Malmo~\cite{johnson_malmo_2016},
SC2LE~\cite{vinyals_starcraft_2017}, and SMAC~\cite{samvelyan_starcraft_2019})
without necessarily grounding in practical applications.

In addition, we argue that securing an autonomous network defender will need
innovation not just in the learning and decision-making algorithms (e.g., to
make them more robust against poisoning and evasion attacks), but also, it will
require the integration of multiple approaches aimed at minimizing the
probability of invalid behavior. Concretely, our work aims to be amenable to the
development of approaches, based on formal methods, to ensure that actions of a
network defender do not cause unintended behavior; e.g., by ensuring that
updates (by an autonomous blue agent) to packet forwarding policies do not
violate packet-traversal goals aimed at segregating flow of information in a
network or preventing distributed denial of service attacks. 

\subsection*{Contributions}

This paper highlights two main contributions: (i) a novel approach
for network environment design to apply RL for network defense; and (ii) a
framework that implements our network environment design concepts to enable
RL and cybersecurity researchers.

\textbf{(i) Network environment design for autonomous network
defense:} Our work proposes aspects of network models that must be available to
researchers seeking to apply RL to reconfigure networks to mitigate
cyberattacks. While network environment design is one task that directly
addresses the problem of enabling progress toward learning a complex task, our
work addresses the more fundamental problem that network defense cannot be
defined as a single game with a simple set of rules. This aspect is especially
important when applying RL to solving a problem that is expected to change over
time. 
Autonomous network defenders must not only be able to reconfigure hosts
and networks to be able to mitigate common adversaries, such as those described
in MITRE's ATT\&CK framework~\cite{strom_mitre_2018}, defenders must
increasingly be concerned about more sophisticated adversaries that target
autonomous cyber defense. Our work takes a novel approach for modeling emerging
threats via state-of-the-art probabilistic computing concepts, some implemented
using~\cite{cusumano-towner_gen:_2019}. Section~\ref{sec:evaluation} illustrates
the perils of training an autonomous defender without considering an adversary
capable of deception. Here, an adversary or attacker would perform deception
using adversarial machine learning approaches such as poisoning and evasion
attacks to manipulate the RL agent defender for the benefit of the adversary.
This is in contrast to a defender employing cyber-deception capabilities such as
honeypots or honeynets to deceive and confuse an attacker. We discuss how, in
order to evaluate the robustness of a RL network defender against deception
attacks, the learning framework must enable the modeling of attackers that
perform evasion and poisoning attacks through \emph{indirect manipulation} of
observations. This contrasts with prior work that has assumed that observations
(e.g., images) can be directly manipulated by an adversary. Such direct
manipulation of observations in network defense involves high degrees of log and
traffic manipulation that correspond to an adversary that would be too strong to
be seeking to manipulate the AI-component of the defender, as an initial attack
vector. As the other non-AI system vulnerabilities become more secure and
trusted, our research can evolve to address deception attacks from sophisticated
adversaries who now pivot to exploit the AI-component of the defender.

\textbf{(ii) FARLAND, a framework to develop and
evaluate approaches for RL-enabled network defense: } We describe the
architecture and an implementation of FARLAND, one of the first software
frameworks specifically designed to enable cyber defense research supported by
AI techniques. Our implementation includes
concrete examples that can be extended by others to develop increasingly more
sophisticated network defenders using RL. These examples help researchers to
gain insights about the following types of questions: 
\begin{itemize}
\item To what extent does FARLAND allow the development of RL agents to defend
networks against adversaries that behave according to a subset of known tactics
and techniques?
\item How does the performance (for training and inference) of a defending agent
change as a result of modifications to the network model and to the assumptions
about normal behavior (by gray agents)? 
\item How susceptible is the performance of such a defending agent to deviations
in behavior by the adversary? 
\item To what extent can an adversary systematically produce behavior that will
influence the behavior of the defending agent in specific ways?
\end{itemize}

The rest of this paper is organized as follows. We start by motivating the need
for a framework like FARLAND, emphasizing
that while dozens of papers have explored the use of RL for
cybersecurity~\cite{nguyen_deep_2019}, the work is difficult to reproduce and
generalize. We highlight how no other system
robustly addresses the problem of network environment design to understand how 
concrete approaches scale and perform as researchers vary network conditions. In particular, we discuss how
there is not a mature understanding about how to protect the
learning and decision-making processes of a RL agent. We then give an overview
of the computing tasks necessary for understanding how to use RL for security
and to study the security aspects of applying RL. We explain why existing
systems without comprehensive network environment design
features 
do not readily support the problems of ``applying RL for cybersecurity"
and ``securing the RL processes". We then give an overview of 
FARLAND's abstractions that
 allow us to model interactions between RL-enabled
network defenders and adversaries from which the defending agent will learn.  We
build on these abstractions to describe FARLAND's architecture, discussing in
detail how it enables the implementation of agent and network models. We
illustrate FARLAND’s concepts via concrete examples of defending and attacking
agents. We conclude with a discussion about how FARLAND can be used for
answering questions about the performance of a learning approach and what
problems need to be addressed to develop RL-enabled network defenders robust
against evasion and poisoning attacks. 

%% file: 06_related_work.tex
\section{Related work}
\label{sec:related_work}

RL is a promising AI approach to support the development of autonomous agents
that must implement complex sequential decision making. Over a period of 5 years
alone (2015-2020), more than 1500 papers about RL have been published
in the top 10 Machine Learning conferences\footnote{The conferences are NurIPS,
ICML, ICLR, AAAI, ICVRA, AAMAS, CVPR, IROS, ICCV, ICAI. Source: Microsoft
Academic~\cite{sinha_overview_2015}.}. Of the many papers that are related to
this work, we only highlight those that relate to the major contributions of our
research. 

We are primarily interested in the problem of applying RL for cybersecurity. On
that subject alone, there are dozens of papers, as Nguyen and Reddi
discuss~\cite{nguyen_deep_2019}. To date, papers in this category primarily
address the \emph{feasibility} problem. Namely, they address the general
question of: To what extent is it possible to use RL learning to automate
complex tasks, such as those necessary for cyber defense? 
However, existing work only addresses simple scenarios 
without offering generalization insights.
Furthermore, there is no
mature research to properly address the question of how RL can be
\emph{securely} applied to solve complex tasks in the presence of adversaries.
Multiple papers have highlighted vulnerabilities of RL to adversarial
manipulation (e.g., \cite{xiao_characterizing_2019, behzadan_faults_2018,
kos_delving_2017,
lin_tactics_2017,han_reinforcement_2018,han_adversarial_2019,gleave_adversarial_2019}).
However, there is still no fundamental understanding about how to secure RL
agents during learning and while making decisions.  
Moreover, many papers that have exposed RL's vulnerabilities to adversarial
manipulation have done so under unrealistic assumptions. For example, some work
assumes that observations and/or rewards can be directly manipulated by
adversaries~\cite{han_reinforcement_2018,kos_delving_2017}. We argue that in the
context of defending a network, such data manipulation requires that an attacker
tampers with traffic and client logs. Such kind of manipulation  can be
prevented via traditional means~--e.g., encryption. Our position is that
indirect observation manipulation (also known as \emph{environment
manipulation}~\cite{xiao_characterizing_2019} or \emph{adversarial
policies}~\cite{gleave_adversarial_2019}) and actuator
manipulation~\cite{behzadan_faults_2018} are more realistic threats. Our work
aims to enable the study of such kinds of attacks in the context of autonomously
defending a network.

While the issue of securing RL agents is clearly important, it is more critical
when RL agents are specifically deployed in adversarial situations (e.g., when
using automation to defend an enterprise network). In particular, existing
experimentation systems do not support research to answer the following types of
questions: To what extent can we use RL to develop agents that learn to perform
security-related tasks? Furthermore, how do we measure the robustness of such
agents to deception (poisoning and evasion) attacks? Previous
work~\cite{baillie_cyborg_2020} attempts to provide a framework to answer the
first type of questions. 
In this paper, we argue that to comprehensively address both
questions, it is necessary to be able to model a diverse set of network
environments and adversaries.



%% file: 07_motivation.tex
\section{FARLAND motivation and overview}

Existing environments, libraries, and frameworks~--e.g.
\cite{brockman_openai_2016,beattie_deepmind_2016,johnson_malmo_2016}~--only
support a subset of the tasks necessary to develop AI systems for security. They
either focus on the development and comparison of RL algorithms for a narrow set
of problems not directly linked to real-world applications; focus on the
development of RL algorithms to solve specific real-world problems in
non-adversarial settings; or are aimed at the evaluation of endpoint detection
and response (EDR) systems and cyber defense teams. Systems in the last category
are not necessarily optimized for autonomization (to support autonomous control
or scalable experience replay).

For example, Gym~\cite{brockman_openai_2016}, a system to develop and compare RL
algorithms, does not currently provide an environment that resembles the kind of
adversarial environment that a network defender would face. Similarly, while
DeepMind Lab~\cite{beattie_deepmind_2016} and Malmo~\cite{johnson_malmo_2016}
facilitate the development of environments with various degrees of complexity,
it would be extremely challenging to map one of these environments to an
environment that adequately models the task of defending a network against
sophisticated adversaries. This is also the case for other systems, such as
SC2LE~\cite{vinyals_starcraft_2017} and SMAC~\cite{samvelyan_starcraft_2019},
which support environments with various degrees of partial observability and
multiplayer interactions. As we describe in this section, these kinds of
environments are not suitable for learning to defend networks, especially when
deception (poisoning and evasion attacks) may be a concern. 

Besides environments and systems that aim to improve the state-of-the-art of RL
algorithms, there are other systems that aim to demonstrate the suitability of
RL to solve networking problems, such as AuTO~\cite{chen_auto:_2018} and
Iroko~\cite{ruffy_iroko:_2018}. However, these do not facilitate answering
security research questions because they do not facilitate the modeling of
adversaries. 

Systems to facilitate interactions between attacker (red) and defender (blue)
teams are designed with different goals (e.g., to evaluate EDR systems) from
those that we outline in this section. FARLAND prioritizes support for
performing large numbers of fast simulations and emulations via interfaces that
are suitable for developing AI agents. Systems such as MITRE's Blue vs Red Agent
Wargame eValuation (BRAWL) are designed to be more broadly applicable, but are
not primarily designed for defender autonomization. For example, BRAWL supports
multiple teams of humans competing in a cybergame using command line interfaces
and graphical user interfaces on a Windows operating system. However, BRAWL does
not provide an API to allow an autonomous blue agent to reconfigure networks to
implement protection mechanisms. BRAWL, also does not provide a way to simulate
a large number of experiences without emulating them.

There is prior work that explored the use of RL for cyber
defense~\cite{han_reinforcement_2018, han_adversarial_2019}. However, 
the results
are not generalizable because their assumptions differ significantly from
real-world scenarios. Finally, CybORG~\cite{baillie_cyborg_2020}, like FARLAND,
provides a framework to allow researchers to vary network models and
adversaries; and, it recognizes that a compatible simulation-emulation approach
is needed to enable learning while facilitating model validation. However, the
current CybORG design does not explicitly allow modeling asymmetrical behavior
and goals between the red and blue agents. Also, CybORG currently does not
include adversarial influence, such as deceptive red agent behavior, as an
attack type that must be taken into consideration when investigating how to use
RL in security applications.
Furthermore, FARLAND's use of generative programs to specify
aspects of a network environment allow for the development of unsupervised 
strategies to gradually increase the fidelity of models. CybORG does not offer
similar functionality.

In addition,
FARLAND's architecture differs from CybORG's, in several key aspects: CybORG
models only cover what we call \emph{traditional} red
agents~(\cref{sub:red_agents}), 
whereas
FARLAND models include \emph{traditional} and \emph{deceptive} red agents, as well as,
gray agents with probabilistic behaviors. In addition, CybORG's models use
finite state machines as opposed to FARLAND's richer probabilistic models via
generative programs~\cite{cusumano-towner_gen:_2019}. FARLAND's architecture
is designed to scale emulation tasks, as well as, learning tasks, by leveraging
RLLib~\cite{moritz_ray:_2018}. CybORG does introduce an important advance in
research that applies RL to cybersecurity; however, its current design does not
include some important considerations addressed by FARLAND.

\subsection{Abstractions}
\label{subsec:abstractions}

As with other RL
frameworks~\cite{bellemare_arcade_2012,beattie_deepmind_2016,johnson_malmo_2016},
researchers need adequate abstractions to define \emph{environments}, agents'
actions, observation spaces, and algorithms that agents use to learn how to act.
However, in the case of 
network defense, environment design includes defining assumptions about red and gray agent behavior.  

In FARLAND, many of these aspects are defined via generative
programs~\cite{cusumano-towner_gen:_2019}. Generative  programs define
probability distributions on execution traces. That is, instead of defining
deterministic mappings between inputs and outputs, they define a weighted set
$\{(x, \xi)\}$ of possible execution traces, where $x$ denotes a trace, and
$\xi$ denotes its weight, associated with the probability that running the
corresponding program $\P$ with specific arguments $\alpha$, results in $x$. 

Generative programs allow
researchers to model stochastic aspects of network behavior, such as gray agents
or failures on a network. Additionally, they allow the specification of
inference queries that may be used by a red agent (for deception purposes), or
by a blue agent to quantify uncertainty. More generally, some of these aspects
can also be defined via probabilistic models derived from data (e.g., large
datasets of regular network traffic and logs) provided that these models can be
sampled.

Thus, the process of developing a RL network defender requires
two tasks: \textbf{RL algorithm design} and \textbf{network environment design.} RL
algorithm design tasks are related to specifying how a learning agent will
learn a suitable policy from a concrete collection of action spaces, observation
spaces, and reward functions. RL algorithms for cyberdefense also need to
integrate mechanisms to prevent adversarial manipulation via poisoning and
evasion attacks. Network environment design tasks include the definition of 
action spaces, observation spaces, and reward functions, as well as the
definition of probabilistic models that characterize network dynamics (including how devices are
connected) and the behaviors of red and gray agents.

\subsection{RL for network defense}

RL agents are autonomous systems that learn to execute complex tasks through
performing actions on an environment and observing the effects of their actions
on this environment. Observations consist of state features and rewards. Rewards
are quantities that encode the desirability of the effects. The goal of a
RL-agent is to learn to act to maximize the sum of expected rewards over a
period of time. Some aspects of engineering reward functions and observation
spaces in the context of cybersecurity are analogous to efforts to design other
RL-enabled systems. 
However, there are key differences when applying RL for cybersecurity, which we
illustrate in Section~\ref{sec:developing_agents}. 

Implementing this type of RL agent involves three types of computations (see
Figure 1): serving, training, and simulation (or emulation\footnote{As we
describe in Section 3, in this paper, we refer to emulation when implementing
agents that perform actions on virtual networks with containerized hosts. We
refer to simulation when agents perform actions on modeled networks.})
(cf.~\cite{moritz_ray:_2018}, for example). Serving computations involve the
selection of actions that maximize long-term rewards. Training computations
typically involve solving computationally expensive optimization problems—e.g.
stochastic gradient descent—aimed at improving the current policy (the mapping
to select actions when the agent is at a specific state). Finally, simulation or
emulation computations can be a broad class of computations aimed at generating
experiences for the RL agent. 

In the context of learning to defend a network, the training and
simulation/emulation tasks are both computationally intensive. Emulation
involves the creation of virtual networks with virtual hosts running real
applications on real application systems. Hence, emulation will require
computation proportional to the size of the network and the complexity of the
applications that are emulated. In addition, simulation actually involves the
computations of red agents, which may include searches and solving optimization
problems to decide how and when to create noisy observations. Finally, training
tasks also require a large number of operations on GPUs. 

FARLAND builds on RLLib's abstractions to be able to easily add resources as
they are available without having to specifically redesign algorithms or
distributed computation details every time that a researcher wishes to add more
CPUs, GPUs, memory, etc.

\subsection{Network dynamics models}
\label{subsec:network_dynamics}

FARLAND supports a simulation mode and an emulation mode. Emulation requires a
virtualization environment that supports the creation of $n \cdot m$ containers,
where $n$ is the number of hosts on the network, and $m$ is the number of
parallel emulations that will be running during training.

As we illustrate in Figure 2, a typical FARLAND deployment requires a
high-performance computing system with GPUs to support the agents’ computing
tasks. The computing tasks to support emulation do not require GPUs. However,
these do require a large number of CPUs. Furthermore, because emulated hosts
execute adversarial programs with high privileges, typical FARLAND deployments
may require an additional layer of separation. 

A separate aspect related to simulating normal conditions on a network is the
development of gray agents. Gray agents execute actions on the network according
to probabilistic models. The types of computations necessary to implement gray
agents’ behavior will largely depend on the complexity of the probabilistic
models.

\begin{figure}[H]
    \centering
      \includegraphics[width=.50\textwidth]{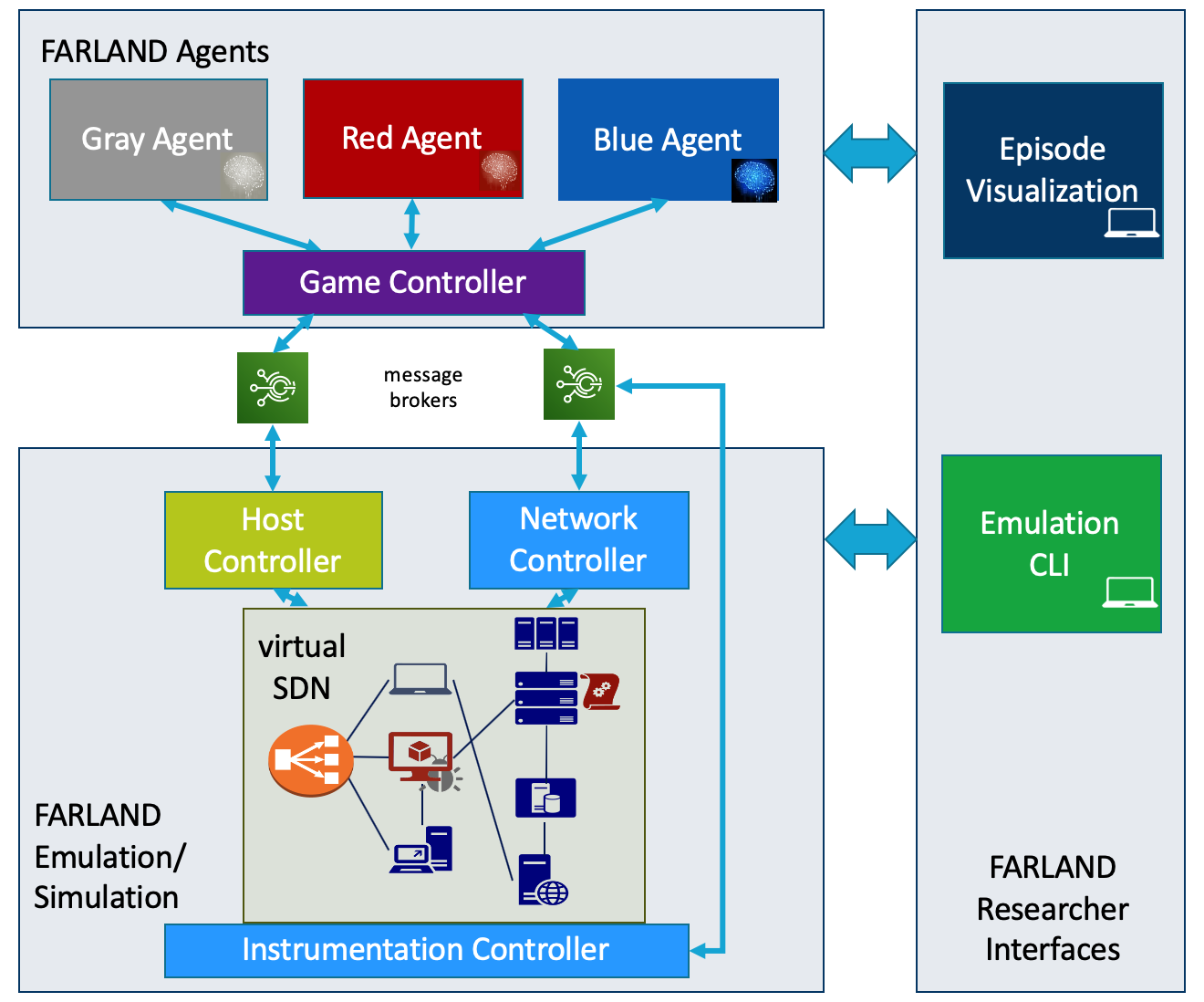}
    \caption[Dos]{Overview of FARLAND components. Agents typically run on a GPU supported system. Emulation runs on a system with an additional virtualization layer to limit the effects of running adversarial programs on shared HPCs. User interfaces run on common desktops.}
    \label{fig:img_tags}
\end{figure}

\subsection{Probabilistic adversarial models}
\label{subsec:adv_models}

FARLAND aims to enable RL to solve network security problems and defend against
sophisticated adversaries. This includes the development and modeling of
adversaries with traditional tactics, techniques, and procedures, as well as
adversaries with deception capabilities. Ultimately, FARLAND aims to evaluate
the robustness of autonomous defenders.

FARLAND simulates interactions of three types of agents with a network. Gray
agents behave according to a probabilistic model that characterizes the expected
behavior of all authorized users in the network. Red agents interact with the
network according to well-known tactics and techniques characterized by MITRE's
ATT\&CK Framework~\cite{strom_mitre_2018}; or, they perform actions on the
network with the goal of deceiving a RL-enabled blue agent. A blue agent
reconfigures the network to mitigate attacks while maintaining acceptable levels
of service. 

To inform the design and evaluation of defenses for robust RL-enabled agents,
FARLAND enables the development of sophisticated adversaries. While blue and red
agents interact with the same environment, they do so in fundamentally different
ways. Generally, network defenders' actions involve network and host
reconfigurations. In contrast, adversaries' actions correspond to actions that
compromised hosts perform on the network. 

Under some threat models, adversaries' intelligence can be modeled via
planning-based intelligence. However, we are interested in modeling adversaries
that purposely manipulate observations in order to influence the blue agent’s
behavior. As we describe in subsequent sections, such a sophisticated attacker
requires computations to predict a set of actions (e.g., via probabilistic
inference) on the network that are likely to result in a targeted observation
vector. The estimation of such a targeted observation vector correspondingly
involves the computation of a solution to an optimization problem.

%% file: 08_game_abstraction.tex
\section{FARLAND game}
\label{sec:praxis_game}

As it is common, when developing RL agents, we use a game abstraction to frame
the learning task. The goal of a cyber defender agent is to learn to achieve a
high score by performing actions on a network to maintain an acceptable level of
service for authorized hosts subject to a resource budget and while preventing
unauthorized access. In other words, the FARLAND game is an abstraction where
the learning agent has to master an operator role in a cybersecurity operation
center (CSOC). 

Network environment models define the dynamics of a game. In other words, a
FARLAND game starts with a preconfigured virtual software-defined network
(SDN) consisting of hosts, switches, network functions, and a network
controller. During the course of a game play, an adversary compromises some
hosts by using a subset of the techniques described in ATT\&CK or by performing
a set of deceiving actions. The defender performs actions on the network to
mitigate the effects of attacks and to maintain the network operating as
intended. A game ends after one of the following: when the attacker achieves (or
believes that it has achieved) its high-level goals; when the blue agent
prematurely terminates the game (with an associated penalty); when the red agent
concedes the game as a result of being contained by the blue agent; or after a
timeout.

FARLAND episodes can be described via sequences of agent actions and states,
which correspond to the application state of agents (red, gray, and blue); and
the state of the network---i.e., hosts, files, databases, and communications. 

A network environment model includes a model of the network,
but, more importantly, it includes a set of probabilistic models in the form of
generative programs that define the behavior of red and gray agents. In
addition, it includes definitions of action and observation spaces for the blue
agent. We describe these three types of agents next.

\subsection{Red agents}
\label{sub:red_agents}

The network will be attacked by one red agent, which controls the behavior of
multiple remote administration tool (RAT) agents. RATs are applications that run
on compromised hosts in order to execute techniques orchestrated remotely by the
master red agent.  There are two kinds of red agents: \emph{(traditional)} one
that aims to achieve one high-level goal, consistent with the ATT\&CK framework
(e.g., to exfiltrate data from a Crown Jewels Server (CJS); to interrupt service
by performing a distributed denial of service attack on a messaging broker; or,
to create an unauthorized LDAP user to enable a future attack); and another one
\emph{(deceptive)} that aims to deceive an autonomous blue agent to decrease its
performance.

The adversary achieves these goals via techniques executed by compromised hosts.
During the course of a game, hosts become compromised by implementing strategies
orchestrated by a planning agent or, by any agent whose decision-making is
described by a probabilistic program in Gen~\cite{cusumano-towner_gen:_2019}.
Additionally, for the model to be emulated, there needs to be a corresponding
implementation of each red action that the agent can perform. These typically
correspond to known implementations of techniques in the ATT\&CK framework
(e.g., to perform discovery or lateral movement steps).

Deceptive agents also perform actions that result from variations of techniques
from the ATT\&CK framework (but without following the corresponding tactics) or
additional procedures that resemble actions from gray agents. The goal of this
agent is to deceive the blue agent to (a) decrease its performance; or, (b) to
induce specific actions from the blue agent. Ultimately, the set of actions
available to red agents can be very broad, ranging from a subset of procedures
derived from the ATT\&CK framework, to an arbitrary set of programs that can be
executed by hosts on the network.

\subsection{Gray agents}

As part of modeling normal network behavior, FARLAND also allows the
specification of gray agents with non-adversarial behavior. This is an important
aspect to prevent the RL-enabled agent from simply learning to react to any
network stimuli (corresponding to the actions of a red agent). 

Non-adversarial behavior is implemented by orchestrating several applications
that perform actions on hosts and on the network. For example, these
applications may cause a host to execute a program that touches specific files
on the host, increase its CPU utilization, or generate various types of network
traffic, such as AMQP, HTTP, and SSH traffic. The behavior of these applications
is specified using probabilistic programs that reflect the stochastic nature of
gray agent events.  Such applications include, for example: an application that
sends (from each host) logging information via AMQP messages; and an application
that generates remote procedure calls (RPCs) from one host to another using AMQP
and REST messages to simulate distributed applications using microservices, IoT
applications, or computationally intensive applications; an application that
transfers data between hosts and executes tasks on other hosts over SSH.  

\subsection{Blue agent}

The blue agent is the RL-enabled agent that defends the network. Thus, this blue
agent also induces game state changes, for example, when it reconfigures the
network.  The defender agent protects the network by performing actions to
contain attacks; to migrate or replicate services; or to deceive the attacker to
either obtain more data to enable attribution, or to identify the purposes of
the attack. 

For example, a defender may isolate one or more hosts by changing packet
processing rules; migrate a service by deploying a new virtual host and changing
the network configuration to redirect future traffic to the new host and
disconnecting the old service; replicate a service by deploying a new virtual
host to add redundancy; reduce connectivity by changing packet processing rules;
change logging policies; or create honey networks by deploying new virtual hosts
and changing the network configuration to connect compromised hosts to fake
hosts and disconnecting compromised hosts from specific regions of the network.
The agent may add or remove filtering policies or add waypoints (forcing packets
to travel through one or multiple network functions before reaching their
destination) in packet trajectories. A blue agent may do nothing, keeping the
network and host configurations intact.

\subsection{Network environment design}

A full CSOC model for FARLAND requires definitions of sets of actions, states,
and observations; as well as, definitions of transition models and goal tests
for red and gray agents. In addition, for blue agents, it is necessary to
specify reward functions. We describe each of these next.

\textbf{Defining agent behavior.} Each of the agents takes actions that alter
the state of the network (i.e., cause the environment to transition from one
state to another). However, each agent only partially and imperfectly observes
the state of the network. 

Hence, defining the behavior of red, gray, and blue agents requires definitions
of \emph{actions} (in $\A$) that affect the network, that agents take based on
\emph{observations} (in $\Omega$), which depend on the state of the network (in
$\S$). Thus, model designers also need to specify \emph{observation filters},
that is, mappings (in $\{ m :\S \rightarrow \Omega \}$) that extract
observations from the network state, from which the agent estimates a state. The
process of determining future actions (the \emph{policy}) is provided via a
probabilistic program in the case of red and gray agents. Blue agents learn
policies via RL algorithms. Different agents may estimate states differently,
taking into account that each has different goals and observability
capabilities. 

\emph{Transition models}, i.e., models that describe the probabilities of
transitioning from one state to another do not need need to be provided when
using FARLAND's emulation (cf. ~\ref{sec:csoc_emulation}). However, to enable
fast learning via simulation, it is necessary to specify these transition
models. Explicitly, agent actions correspond to programs that interact with the
network. Red and gray actions are programs that run on hosts, while blue actions
are programs that may run on hosts, as well as, on controllers that alter
network configurations. When using FARLAND's emulation, these are arbitrary
programs that can be run on virtual systems, and hence the transition from one
state to another is implicitly determined by the programs that are run. When
interactions with the network are simulated, model designers need to specify the
network transformations that would result from taking an action. In either case,
agents need to extract information from the network state to estimate their own
imperfect state. 

For example, a \emph{lateral move} action may be a program that implements a red
agent's procedure to compromise a host from another one. When emulated, the
transition from one network state to another is the result of actually executing
the corresponding program on a virtual network. When the action is simulated,
the transition model needs to describe how the action is expected to affect the
creation of packets and log entries and the presence of a RAT in the target host
after taking the action. The red agent will also need to apply an observation
filter to update its belief state (i.e., the red agent may keep track of the
network hosts that are known and those that are believed to be compromised).  

When actions are emulated, observation filters are specified via parsing
functions of standard streams (stdout and stderr) and exit values. Note that, in
particular, these observation filters help to deal with programs that produce
large amounts of output and with programs that may terminate abruptly (e.g.,
because they may be terminated by a protection mechanism).

\textbf{Game state.} Defining state spaces is another aspect that researchers
must specify. The state of a network depends on how devices are connected and
how network configurations change as a result of agents' actions. Typical
network configurations include a set of hosts, which may change during the
course of a FARLAND episode. Network state is internally represented as graphs
with attributes. The nodes in the network state graph contain hosts, which also
have attributes. 

Red, gray, and blue agents each keep one such graph. These in general are
different because agents have different knowledge about the hosts in the network
and each agent may keep track of different host attributes. For example, blue
agents may have full visibility about which nodes are connected on a network and
how (including switches and network functions); services that are running in
each host and which ports are used; information about which hosts contain the
crown jewels to protect (sensitive files, AMQ brokers, and LDAP servers); and
packet forwarding rules. In contrast, red agents only have partial and imperfect
information about hosts and network configurations. Also, when the blue agent
uses honey networks to learn about an adversary, the blue agent may use
attributes to distinguish fake from real hosts. These attributes may not be
accessible by the red agent. In contrast, a red agent may use host attributes to
distinguish between compromised and non-compromised hosts.

Graph attributes are pieces of state that may be useful to infer QoS or
malicious activity. For example, snapshot statistics that summarize numbers of
packets delivered and in transit, or more generally, quantities that summarize
the successes and failures of authorized communications on the network.

\textbf{Observations.} Because it is impractical to assume that agents will be
able to fully observe the network state, agents act on partial and imperfect
state information. Part of a threat model is to determine what features of the
state are observable by which agent and with what kind of uncertainty. 

Typically, one may assume that gray agents act on information that would be
available to normal authorized users. Blue agents may have access to much more
information, but it may be necessary to only focus on a subset of the state
space for the purposes of making the learning process feasible. For example, a
blue agent may take actions based on an observation matrix that summarizes the
state of the environment at a snapshot in time, summarizing network information
and host activity information primarily related to quality of service and threat
indicators.

Network information may include specific events (e.g., host $A$ sent a file to
host $B$ via scp), or they may include network statistics that, for example,
describe volumes of traffic in network regions or between specific hosts. These
types of information would be typically gathered by network appliances, but in
our case, they can also be gathered by a monitor (FARLAND Instrumentation
Controller) that operates at a lower virtualization level than hosts and network
devices. 

Host activity information describes events observed by monitors on the hosts and
do not necessarily involve network events or generation of traffic. These kinds
of information may be collected by low-level daemons running on hosts, or as in
the case of network information, these can be collected by the FARLAND
Instrumentation Controller.

\textbf{Rewards.} Reward functions are specified by the researcher and take as
inputs features of the network state; costs associated with deploying or
repairing services; and,  potentially subjective representations of how good or
bad certain states are with regard to a set of security goals. These may be
subjective, because different organizations may value security goals
differently, and deciding how much worse an outcome is than another outcome may
not be simple. However, broadly speaking, indications of successful compromises
or service degradation should incur penalties. Successful attack containment and
adequate service would result in positive rewards. For a given environment, the
set of suitable reward functions is not unique.  As prior research has shown, in
general, there can be infinitely many reward functions, which may be consistent
with the desired behavior of the blue agent~\cite{nguyen_inverse_2015}.  
Describing a principled methodology to select a set of reward functions for a
given environment and threat model is an issue that we plan to study in future
work.

%% file: 09_architecture.tex
\section{CSOC emulation}
\label{sec:csoc_emulation}

As we discussed in Section~\ref{sec:praxis_game}, FARLAND game is an abstraction
to facilitate the development of a RL-enabled CSOC operator. However, to ensure
that network and adversarial models are grounded in realistic assumptions, they
should be informed and validated through emulation. This plays an important role
in engineering observation spaces and in understanding the manipulation
capabilities of an adversary. For these reasons, FARLAND was designed with the
idea of providing a single layer to interact with a \emph{CSOC environment} for
the two possible types of environments that a blue agent can interact with~--a
simulated environment and an emulated environment. Interactions with a simulated
environment accelerate the learning process, while interactions with an emulated
environment enrich and validate models and approaches. This section, describes
one way to implement FARLAND's CSOC emulation component depicted in Figure
\ref{fig:img_tags}. 

Agents interact with a CSOC environment through a Game Controller via the same
API, regardless of whether the environment is simulated or emulated. Hence, the
Game Controller receives requests to perform an action on either a simulated or
emulated environment. 

The three main emulation aspects include coordinating network configuration
updates, directing hosts to perform actions, and collecting observations. The
Game Controller mediates actions between the red and gray agents and a Host
Controller, and between the blue agent and a Network Controller, a Host
Controller, and an Instrumentation Controller.

By design, communication between the Game Controller and emulated CSOC
environment is brokered using asynchronous messaging. The reasons are twofold:
emulation may require a large number of CPUs, hence a second system may be
required; moreover, because red agent procedures may actually run adverse code,
it may be a good practice to run CSOC emulations in a separate system, as
opposed to a shared high-performance computing system.

\begin{figure}[ht!]
    \centering
      \includegraphics[width=.70\textwidth]{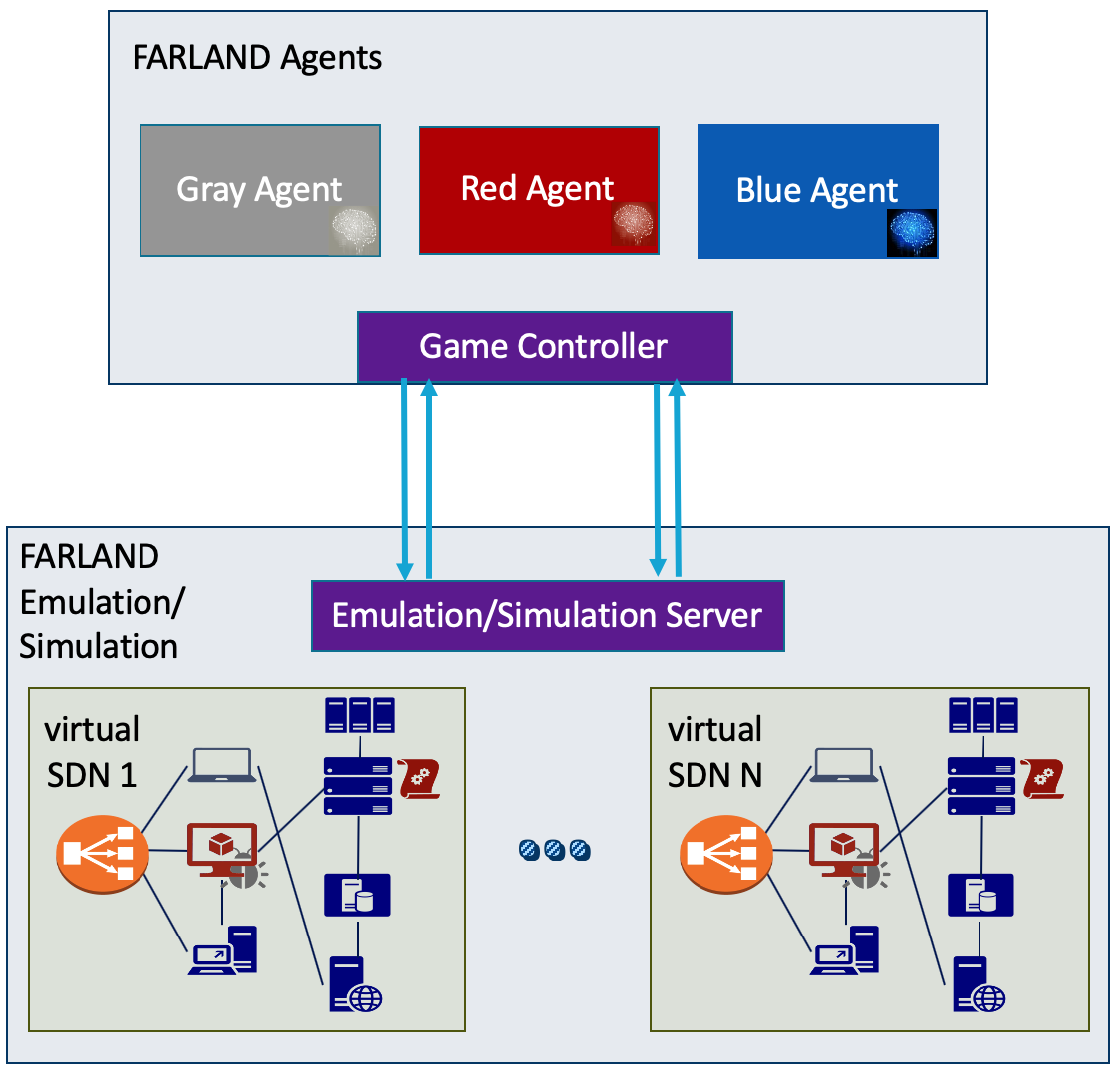}
    \caption[Multiple Emulations]{Training a blue agent involves parallel interactions with multiple instances of a simulated or emulated environment to accelerate learning.}
    \label{fig:img_multiple_emulation}
\end{figure}

In other words, while FARLAND can be run on a single system, we typically run
the emulation component in a separate system. As we illustrate in
Figure~\ref{fig:img_multiple_emulation}, the emulation component will create
multiple virtual networks, depending on the number of RLLib workers that are
selected. RLLib is part of Ray, a distributed framework for AI
applications~\cite{moritz_ray:_2018} that implements common functionality
necessary to implement RL systems~\cite{liang_ray_2017}. 

When emulation is used, the hosts on the network are implemented using Docker
containers and virtual SDNs are created with instances of Containernet.
Containernet is a network virtualization tool that facilitates the
virtualization of SDNs programmatically \cite{peuster_medicine:_2016}.
Containernet was built atop Mininet, a network virtualization tool widely used
by the research community~\cite{lantz_mininet-based_2015}. The Instrumentation
Controller is implemented atop of Falco, an open source monitor developed with
the support of the Cloud Native Computing
Foundation~\cite{noauthor_falco_nodate} and containerized network functions.
Together, these collect data~--stored in a database~--similar to what host-based
and network-based monitoring solutions collect. After a blue agent performs an
action, the Instrumentation Controller gathers observations from the database.
The observation space is configurable and extensible.  

Containers are assumed to be exploitable, so that when a compromised host makes
a lateral move, a host can start a RAT in the target host with elevated
privileges. This allows a compromised host to arbitrarily replace the behavior
of any program in the compromised host. This is necessary to implement deceptive
agents (cf. \cref{sub:deceptive_agents}). Deceptive agents must be able to
generate gray-like actions (e.g., send files via SSH from one host to another),
and they should also be able to frustrate gray actions. The latter is
accomplished by the RAT remapping real programs to fake programs. As a result,
when a gray agent performs an action, the RAT can frustrate the corresponding
observations.    

%% file: 10_developing_agents.tex
\section{Developing agents with FARLAND}
\label{sec:developing_agents}

FARLAND aims to not only support researchers that seek to demonstrate that an
agent can perform security related tasks, but in addition, FARLAND was designed
as a tool to help researchers understand how such learning can be manipulated by
an adversary. Ultimately, FARLAND should also help researchers to evaluate the
performance of defenders to protect the network against traditional behavior
(for example, as described by ATT\&CK’s framework), as well as, to protect their
decision-making policies against deception attacks.

This section does not intend to demonstrate that RL can be effectively used for
implementing robust autonomous cyber defense. Instead, it intends to illustrate
the perils of training an autonomous defender without considering an adversary
capable of deception. This is an important issue because, as soon as RL agents
are deployed, sophisticated attackers will recognize that the decision-making of
these agents can be a new target. Thus, in this section, we illustrate with a
simple example, the kind of adversaries that we can model with FARLAND. Our hope
is that the research community will use FARLAND to develop robust RL mechanisms
against the class of adversaries that we describe. 

\subsection{Deceptive red agents}
\label{sub:deceptive_agents}

As Behzadan and Munir describe~\cite{behzadan_faults_2018}, RL agents can be
manipulated by tampering with sensors, reward signals, actuators, the memory of
the agent, and the environment. However, as research has pointed out (e.g.,
~\cite{xiao_characterizing_2019}), the most concerning and realistic attacks are
those that do not assume that the adversary can directly corrupt data from which
the agent learns. For example, early attacks on RL algorithms
(e.g.,~\cite{kos_delving_2017}), add noise to images from which the agent
learns. This type of direct tampering with observations is not realistic in our
application. Observations from which a network defender will learn are derived
from logs, databases, and network data, whose integrity can be protected by
other means (e.g., using cryptographic approaches). Prior work using RL for
defending SDN networks has assumed that an adversary can directly manipulate
reward signals~\cite{han_reinforcement_2018} and/or
observations~\cite{han_adversarial_2019}.

We argue that one of the most concerning threats may involve \emph{indirect}
manipulation of observations by having a red agent that performs actions on the
network to bias observations, but without directly adding arbitrary noise to an
observation matrix. Similarly, a sophisticated adversary may bias a defender
when the actions of a defender are frustrated. This paper does not illustrate
attacks where an adversary systematically frustrates blue actions; however
FARLAND facilitates the modeling of that type of adversary as well.

We describe an approach for implementing red agents that
indirectly manipulate observations by transforming the environment (the network)
in ways that are assumed possible under an adversarial threat. 

Concretely, the goal of such an adversary is to perform actions on compromised
hosts to generate observations $\hat{o}$, to cause the learning agent to update
its policy $\pi(a|\hat{o}; \theta)$ maximizing the adversarial goals. Thus, the
goal of the adversary is to bias the policy to maximize
$\E_{\pi(a|\hat{o})}[\sum_t \gamma^t \hat{r}_t]$, where $\hat{r}_t$ are rewards
that reflect the adversarial goal, and $\gamma$ is a discounting factor. 

The adversary needs to solve two problems: The first is to decide what $\hat{o}$
would bias the learning agent optimally. The second problem is to determine,
given a target $\hat{o}$, the set of red actions that are most likely to produce
$\hat{o}$. Solving the first problem can be done, for example, by adapting ideas
described by Kos and Song~\cite{kos_delving_2017}. Here, we describe a framework
to solve the second problem using probabilistic programming. 

Probabilistic programs not only allow the implementation of programs with
stochastic behavior, but under appropriate semantics, also facilitate the
estimation of the posterior distribution $p(x|y; \P, \alpha)$ of execution
traces $x$, given that $y$ was observed, subject to a program $\P$ with
parameters $\alpha$. Our key point is that solving the second problem posted
above corresponds to approximating one such posterior distribution. Namely,
given a target observation $\hat{o}$ an adversary would like to know what
execution traces are most likely to have produced $\hat{o}$.

Probabilistic programming languages and systems, such as
Gen~\cite{cusumano-towner_gen:_2019}, facilitate the computation of an
approximate answer to this question $q(x;\P,\alpha, \hat{o}) \approx
p(x|\hat{o}; \P, \alpha)$. This allows an attacker to find a numeric $\hat{o}$
that would optimally bias the blue agent policy, and then find the red actions
that would most likely result in such target $\hat{o}$.

While investigating the full implications of our approach is left for future
work, our hope is that our system, our ideas, and the example illustrated in
\cref{fig:attacks} foster research developing defenses to protect against this
kind of attack.

\subsection{Blue agents against known behavior}
\label{subsec:learningwithknown}

In~\cref{sec:evaluation}, we want to illustrate that, when we train a blue agent
using known TTPs without accounting for potential (poisoning or evasion) attacks
targeting the learning algorithm, its performance can not be easily guaranteed.
To do so, we start by training, using FARLAND, a blue agent against a red agent
that is only assumed to behave in a manner consistent with a subset of behaviors
described in the ATT\&CK framework.  We then show that when we perturb the
behavior of the agent using only gray-like actions, the performance of the blue
agent is significantly degraded.  

Concretely, consider a red agent that is assumed to perform behavior according
to a high-level goals, such as, an exfiltration; the creation of a bogus account
on an LDAP server; or attacking an AMPQ broker through a distributed denial of
service. This red agent will not be aware of what mitigation and deception
techniques are being implemented in the network. The computation and
communication capabilities of the attacker result from the inherent limitations
of the emulated environment. Furthermore, we assume that the attacker is not
able to compromise the SDN Controller, the network functions, switches, or any
other infrastructure that facilitates network configuration and management.

As we discussed, this red agent operates under partial observability. The red
agent starts with limited knowledge about the network configuration, the
services running on hosts, open ports, etc. The policy that the red agent
follows in this case is derived from a planning approach. The red agent
maintains a model of the network configuration (as it is discovered) and host
information, such as which hosts have RATs, discovered user credentials, etc.
The list of node attributes that the red agent maintains includes: the host OS,
the host OS version, its fully qualified domain name (FQDN), its DNS domain, its
IP, the list of used ports with their state (open or closed), e.g., [(22, open),
(80, open)], a list of services, e.g., [ssh, http, ftp], a list of file paths, a
list of AMQ Brokers being attacked, and a list of tags in the set \{‘RAT’,
‘ElevatedRAT’, ‘AdversaryDiscoveredAMQBroker’,
‘AdversaryDeclaredCompromisedAMQBroker’, ‘AdversaryDiscoveredLDAPServer’,
‘AdversaryDiscoveredSpecialHost’, ‘AdversaryDeclaredCompromisedLDAPServer’,
‘AdversaryDeclaredCompromisedSpecialHost’\}.

The blue agent uses cyber-deception as a mitigation strategy, which includes
actions such as isolating nodes, migrating nodes to a new (honey) network, or
migrating nodes to an existing (honey) network (with a reduced action cost).

%% file: 11_evaluation.tex
\section{Evaluating performance}
\label{sec:evaluation}

\begin{figure*}[t!]
    \centering
\begin{tabular}{ccc}
    \includegraphics[width=.3\textwidth]{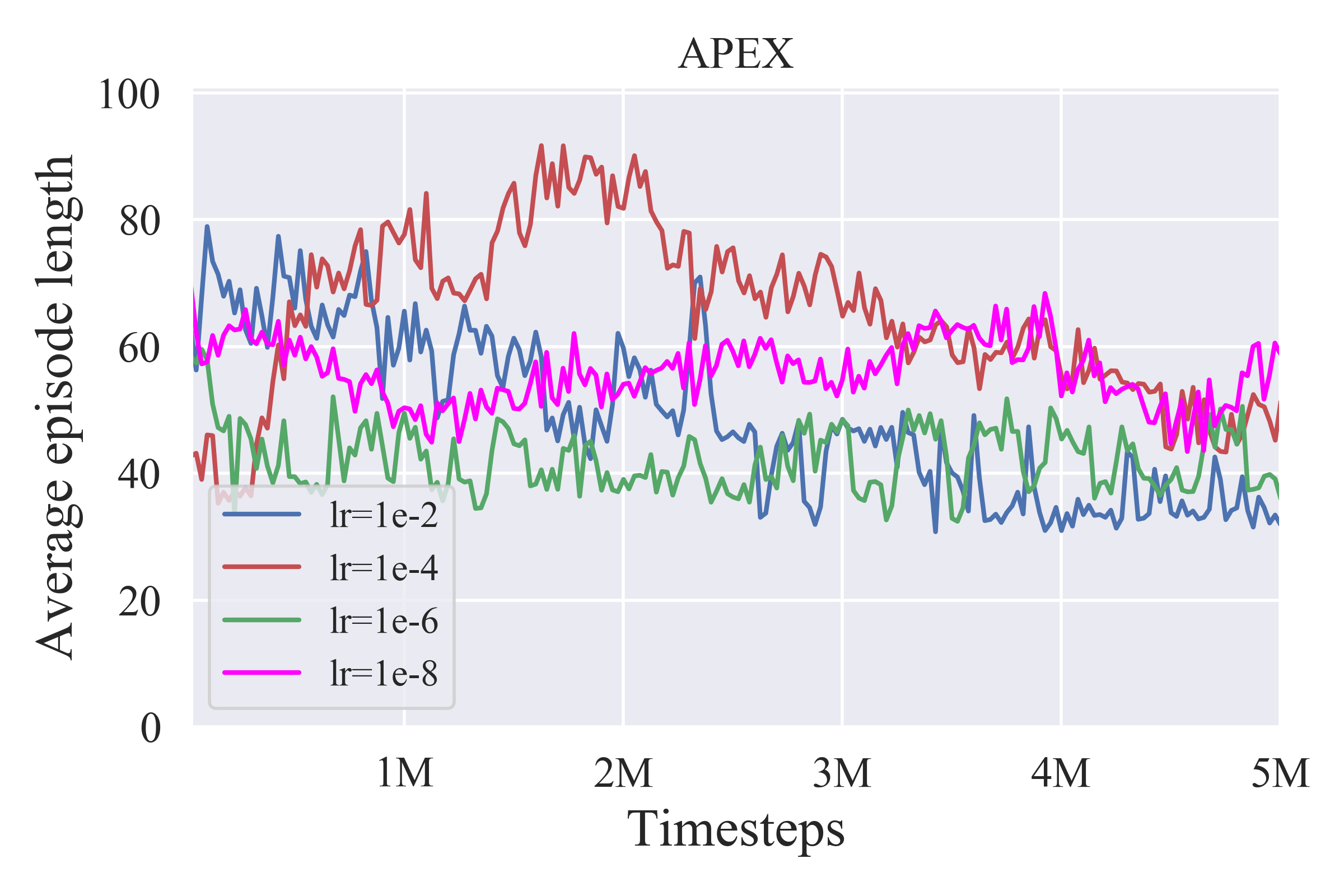} &
    \includegraphics[width=.3\textwidth]{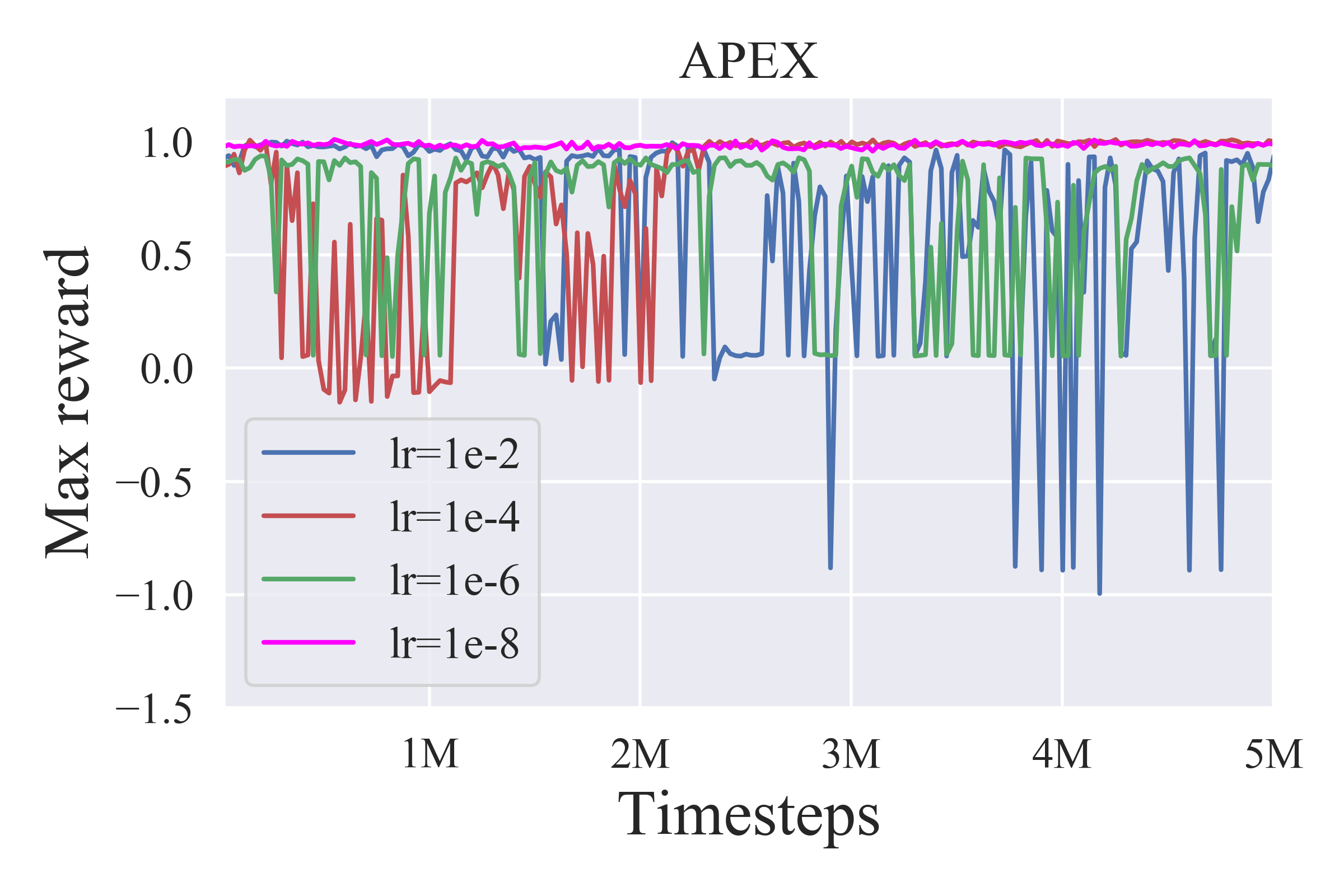} &
    \includegraphics[width=.3\textwidth]{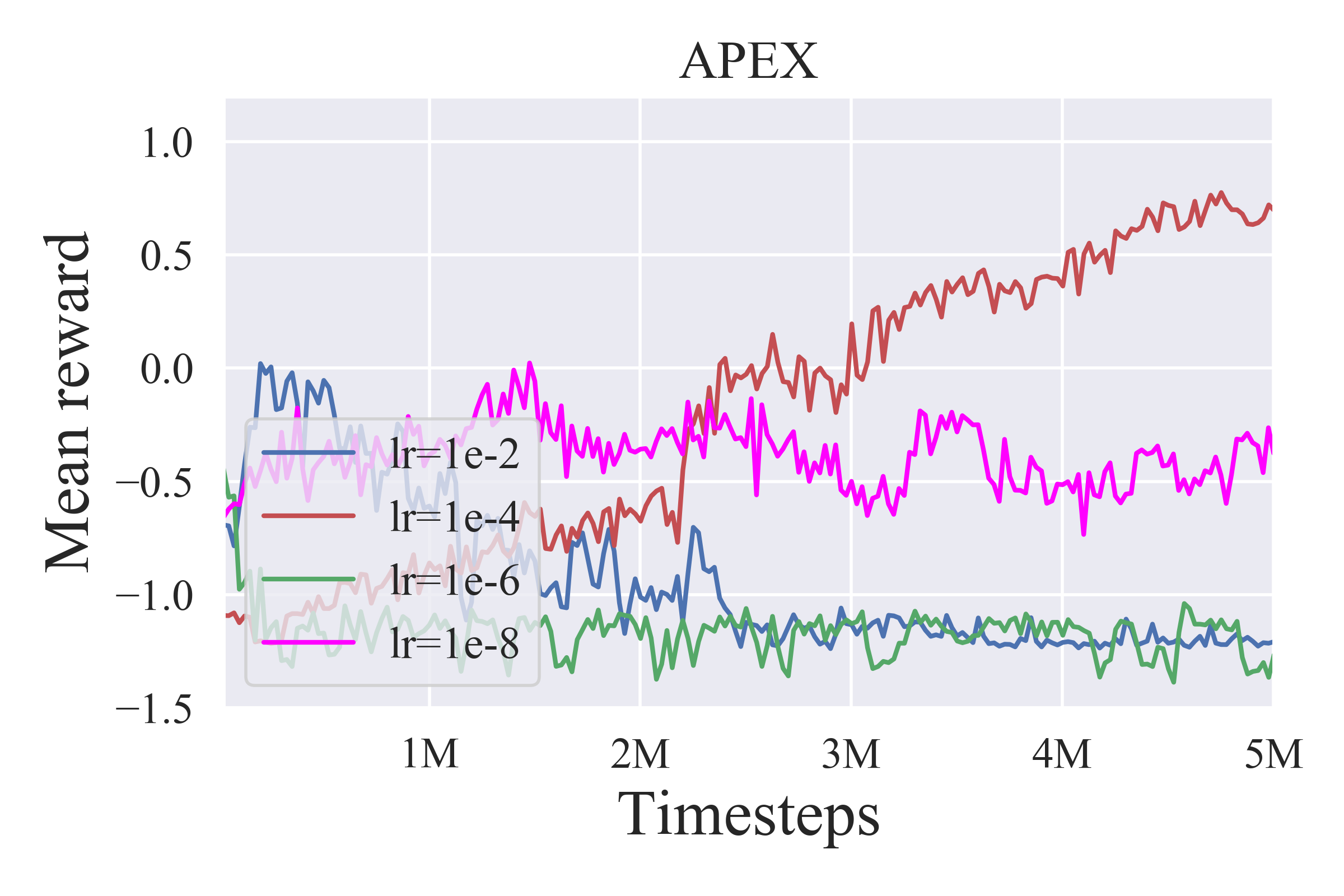} \\
    \includegraphics[width=.3\textwidth]{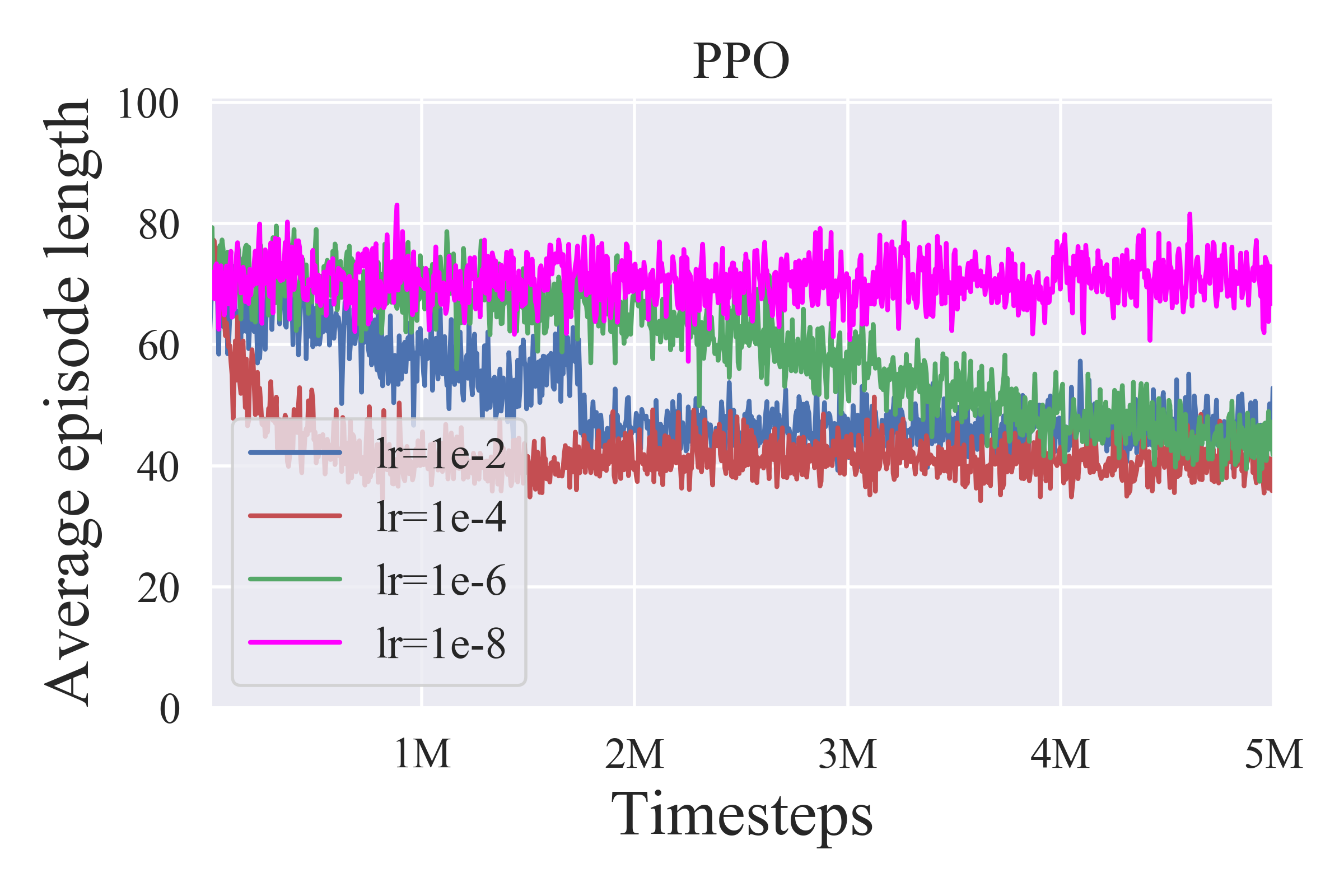} &
    \includegraphics[width=.3\textwidth]{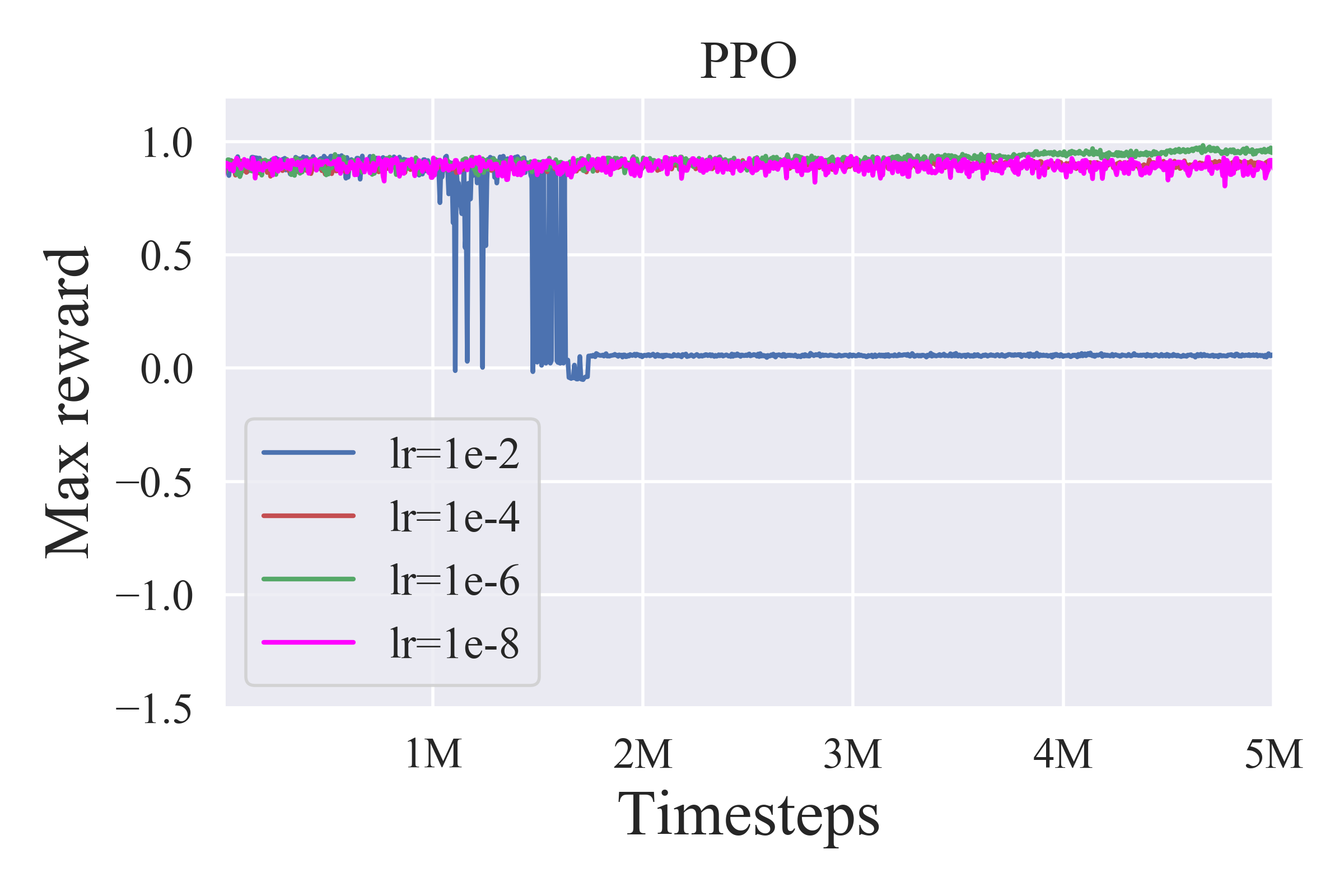} &
    \includegraphics[width=.3\textwidth]{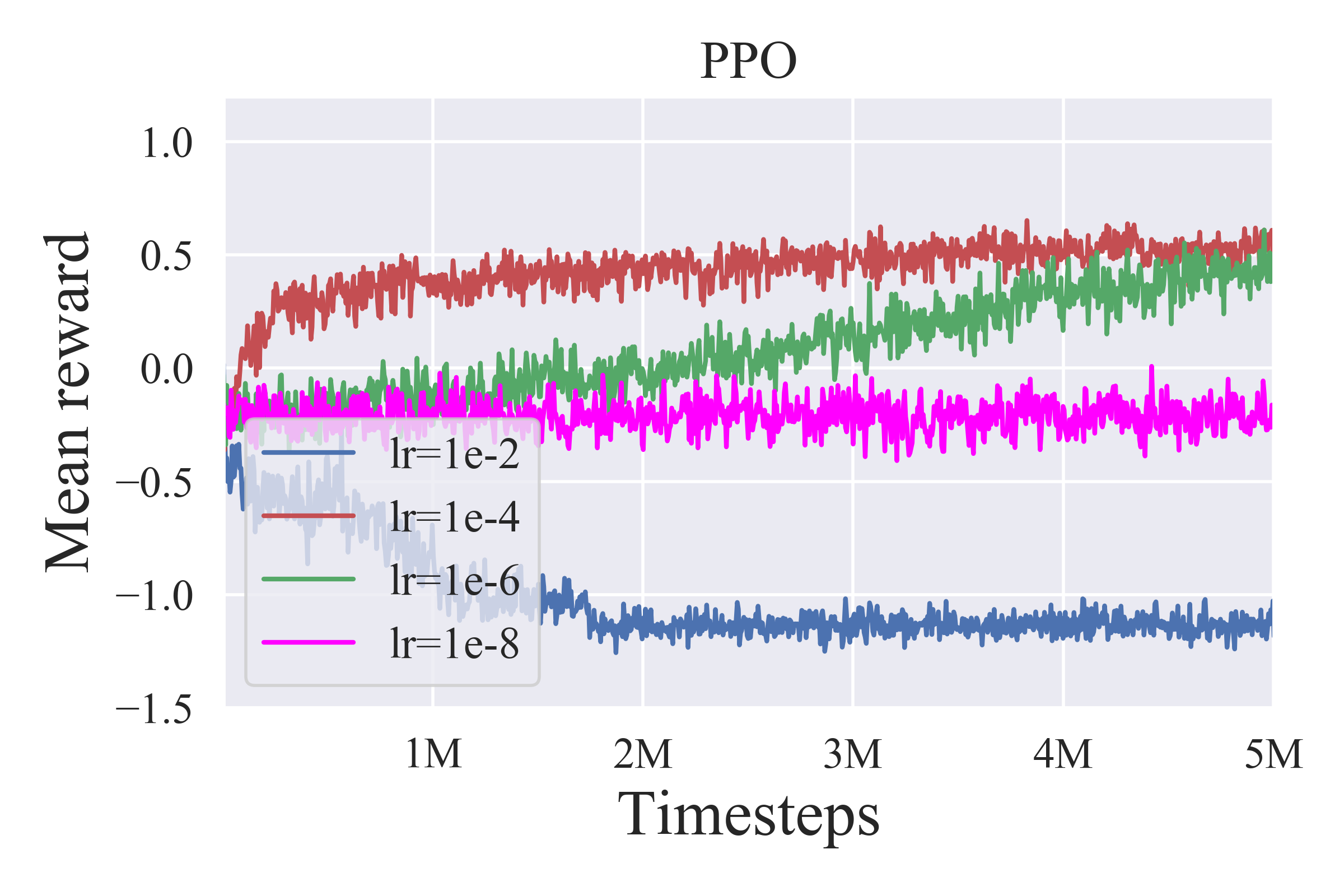} \\
    \includegraphics[width=.3\textwidth]{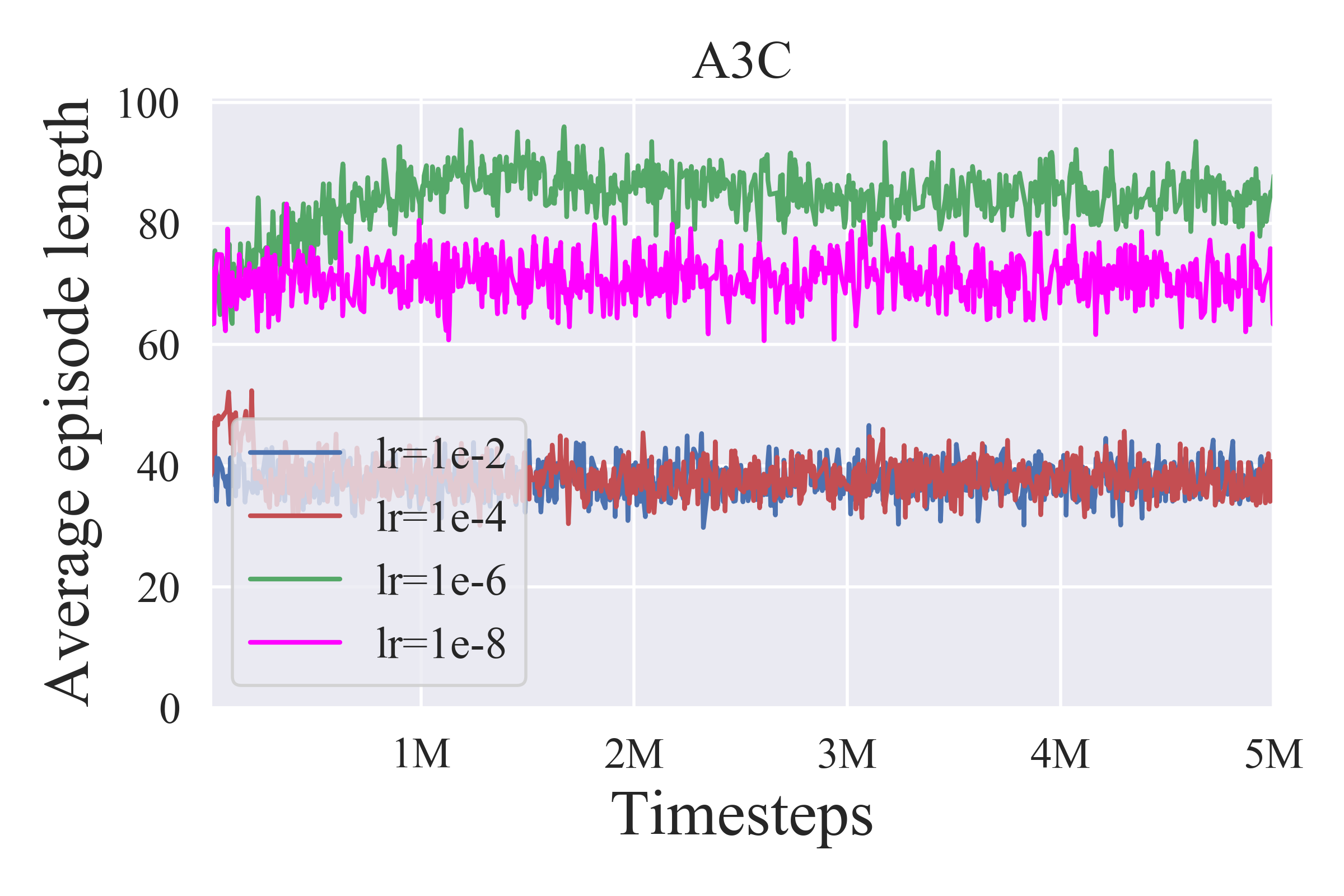} &
    \includegraphics[width=.3\textwidth]{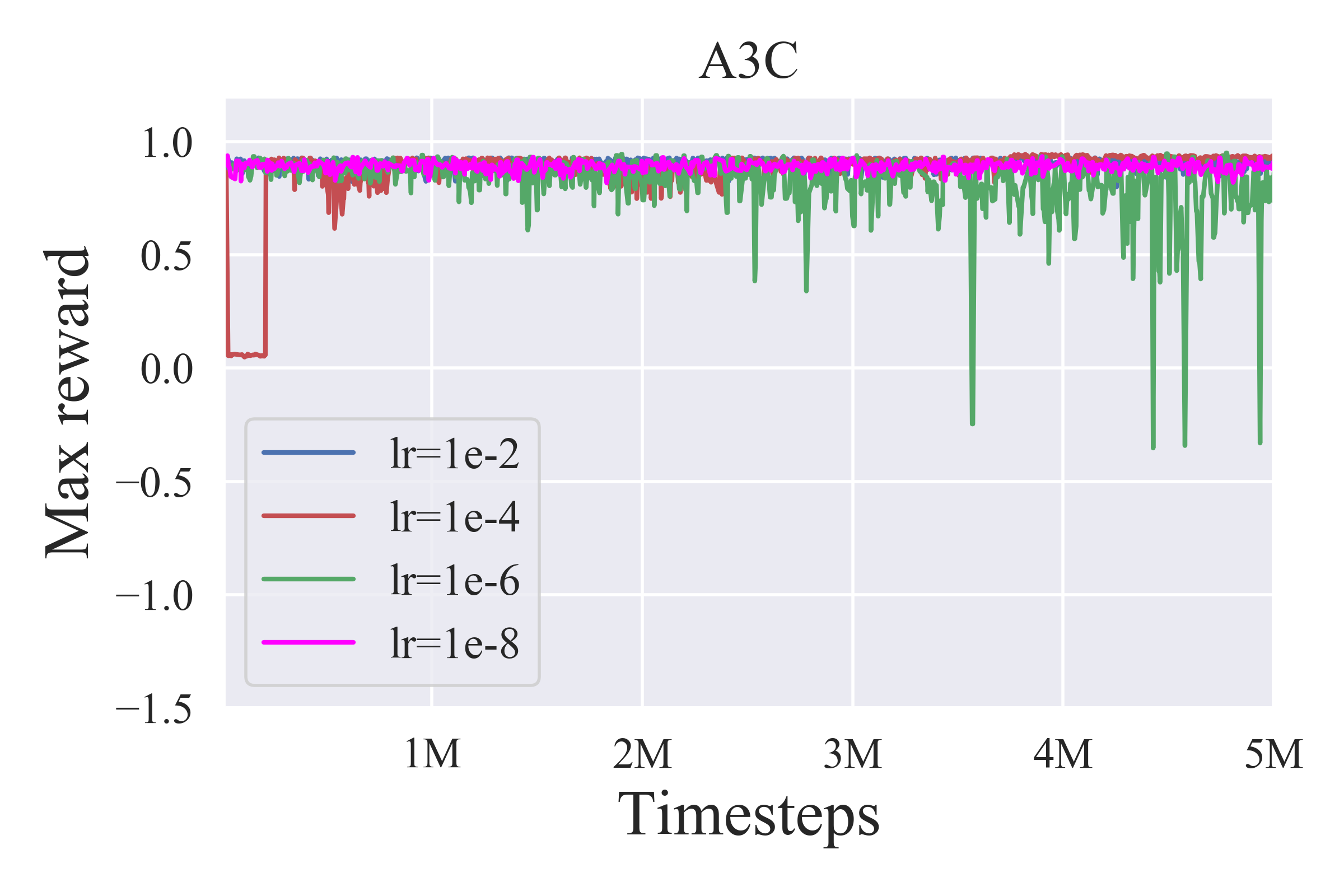} &
    \includegraphics[width=.3\textwidth]{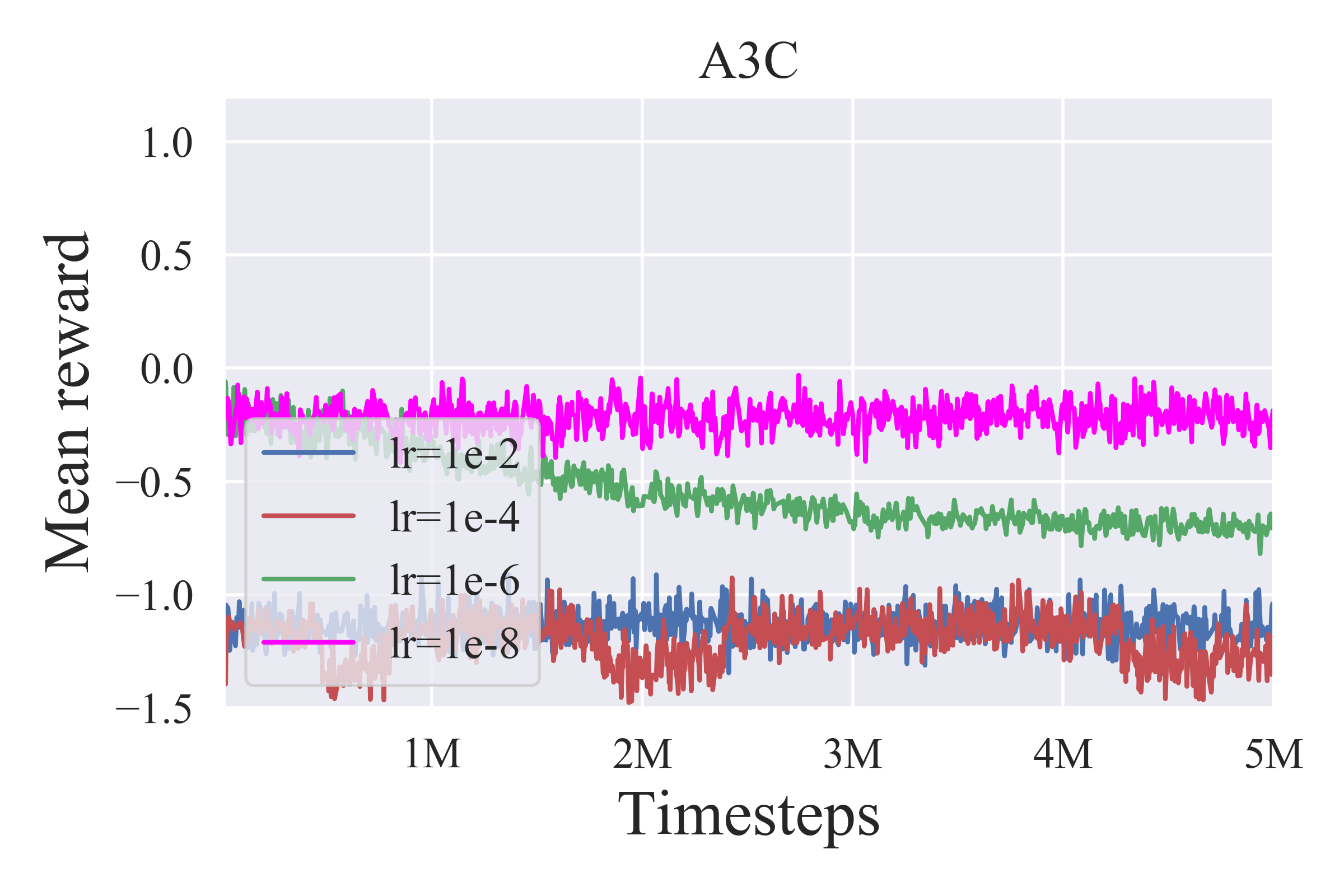} 
\end{tabular}
\caption{Blue agent's performance in a 10-node network against an exfiltration attack (consistent with ATT\&CK) with three different algorithms, APEX-DQN, PPO, and A3C. An episode ends when the red agent exfiltrates a real or a fake crown jewel; or after 100 steps. The left column shows that all three algorithms result in behavior that lets the red agent "win". However some learn to let the red agent win in a fake network rather than in a real network. In the former case, the average reward approaches 1 (e.g., when using APEX-DQN with lr=0.0001). Other configurations, (e.g., PPO with lr=0.01) converge to a policy with poor performance that lets the red agent exfiltrate most of the time. In this case, 5M timesteps correspond to approximately 120K episodes.}
\label{fig:traditional_attack}
\end{figure*}

Section~\ref{subsec:learningwithknown} describes an example representative of
the kind of blue agent that is possible to develop using FARLAND. The
performance of an approach will ultimately depend on the extent to which the
resulting policy produces behavior consistent with expectations. However, the
performance of such agent depends on multiple factors, including reward
functions; action and observation spaces; RL algorithms; and the goals of the
defender. Different security policies would balance QoS and risk differently.
For example, in some cases, the desired behavior of a blue agent is to identify
and isolate compromised hosts as quickly as possible. In other cases, it may be
more valuable to migrate a compromised host to a honey network. 

In the example that we described, we assume that human-level performance entails
(a) detecting and isolating compromised hosts as soon as they exhibit known
malicious behavior; or (b) detecting and migrating compromised hosts to honey
networks as soon as they exhibit malicious behavior. In both cases, a blue agent
would attempt to minimize the disruption of service for uncompromised hosts and
avoid expensive operations (e.g., migrating a host to a honey-network).

Figure ~\ref{fig:traditional_attack} illustrates the performance of a blue agent
that learns with three different algorithms, using various learning rates. The
blue agent experiences an adversary's TTP (exfiltration) consistent with
ATT\&CK, in a 10-node network. The blue agent's action space consists of
selecting a host and choosing whether to isolate it, migrate it to an existing
network, or migrate it to a new network. The reward function favors trapping an
offender in a honey network over isolating it. The reward function also
penalizes unnecessarily isolating or migrating hosts to a honey network when
they are not misbehaving. The agent learns to identify characteristics of
attacks and QoS from the following information: the numbers of SCP events, HTTP
events, AMQ events, SSH events, quiet or aggressive recognizance events
originated at a host; and the numbers of SCP failures, REST request failures,
AMQP failures, SSH failures, and content searches reported by hosts.

FARLAND allows researchers to reason about the following question: How does the
performance (for training and inference) of the blue agent change as a result of
modifications to the network model? Researchers can experiment training agents
with existing algorithms and observe their performance while changing multiple
parameters, including algorithm parameters, such as the algorithm and learning
rates, as well as network configuration parameters. For example, researchers can
change the number of hosts on a network, the network topologies, and the
complexity of the assumed normal behavior (by gray agents).

\begin{figure*}[t!]
    \centering
\begin{tabular}{ccc}
    \includegraphics[width=.3\textwidth]{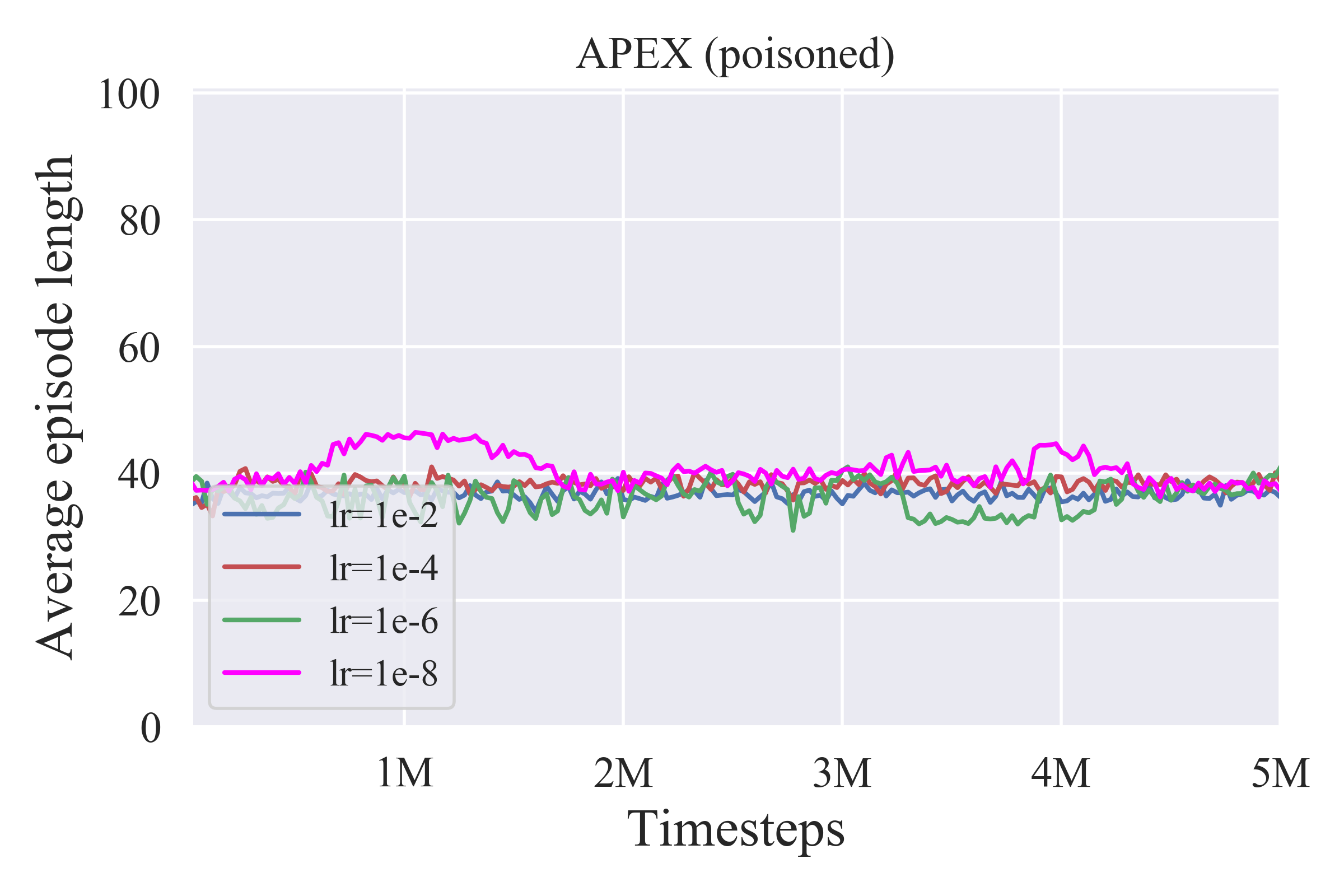} &
    \includegraphics[width=.3\textwidth]{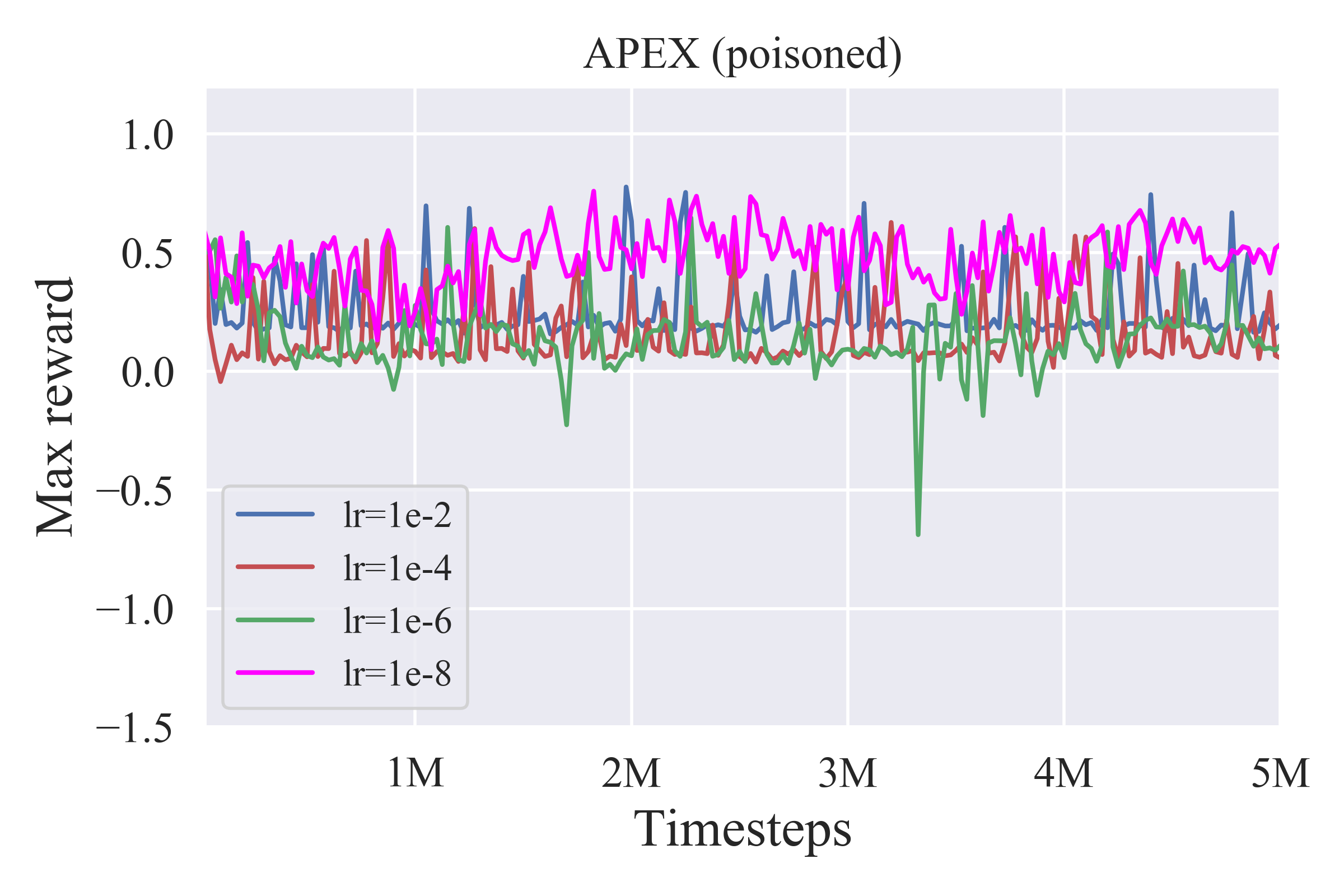} &
    \includegraphics[width=.3\textwidth]{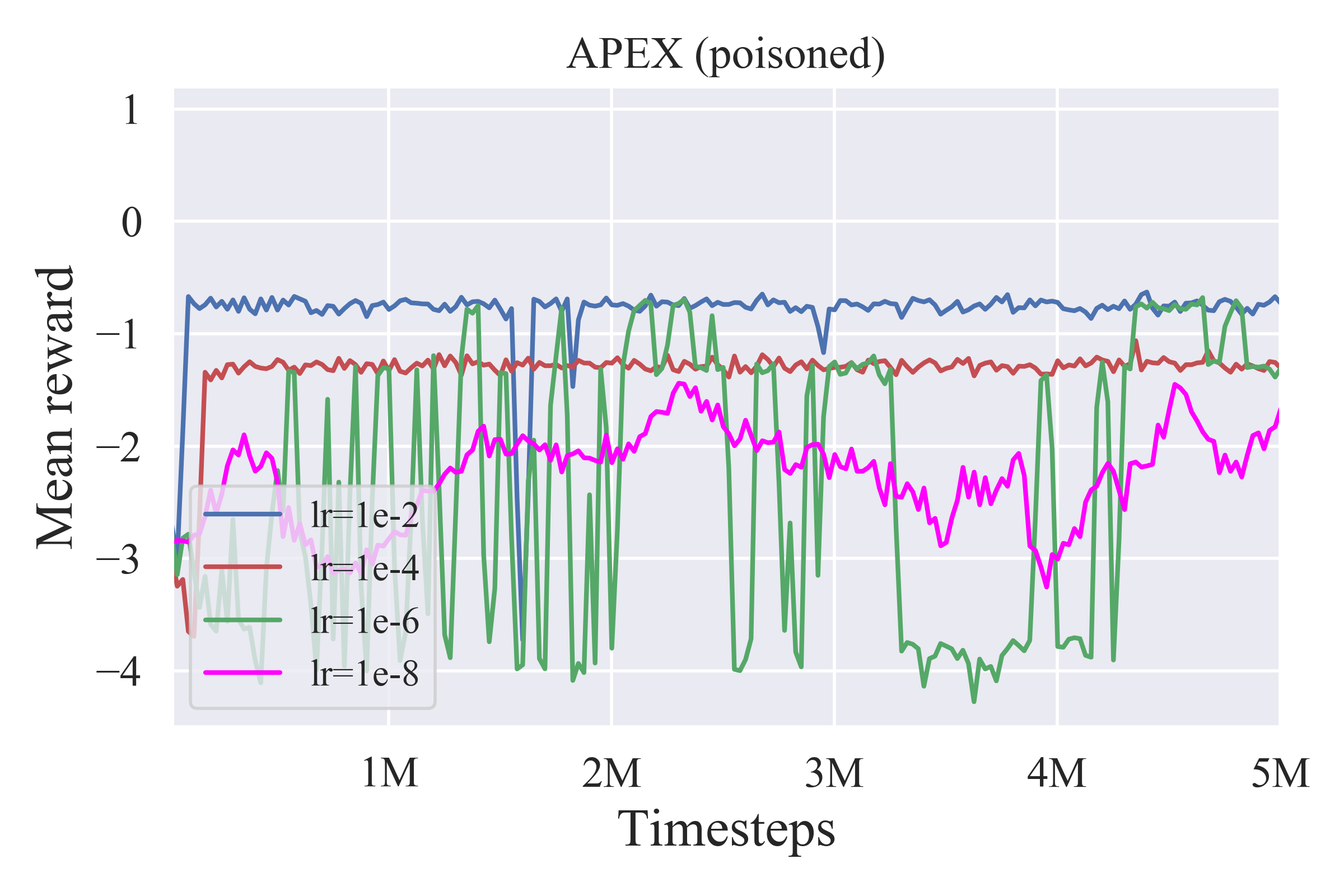} 
\end{tabular}
\caption{Blue agent's performance, learning with APEX-DQN with the same configuration as in Figure ~\ref{fig:traditional_attack}, except that the adversary performs an exfiltration attack with deception. The left plot shows that the red agent successfully exfiltrates the crown jewel most of the time. Furthermore the blue agent fails to learn a suitable policy.}
\label{fig:attacks}
\end{figure*}

In addition, researchers can reason about the extent to which their algorithm is
robust against deception attacks. Figure~\ref{fig:attacks} illustrates the
effect of allowing an adversary to deviate its behavior from the assumptions
outlined in ~\cref{subsec:learningwithknown} with only gray-like actions.

\begin{figure*}[t!]
    \centering
\begin{tabular}{cccc}
    \includegraphics[width=.24\textwidth,height=3.4cm]{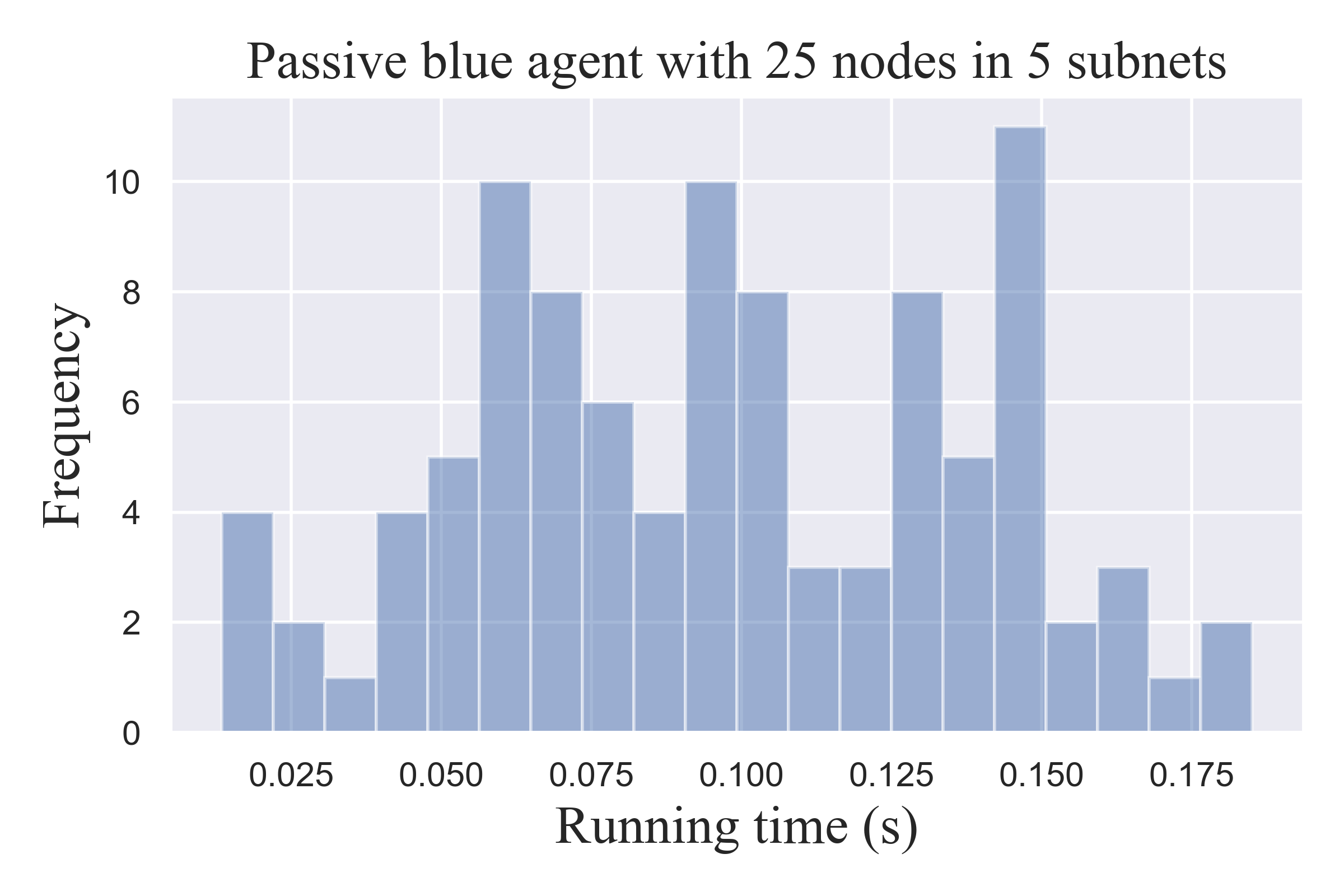} &
    \includegraphics[width=.24\textwidth,height=3.4cm]{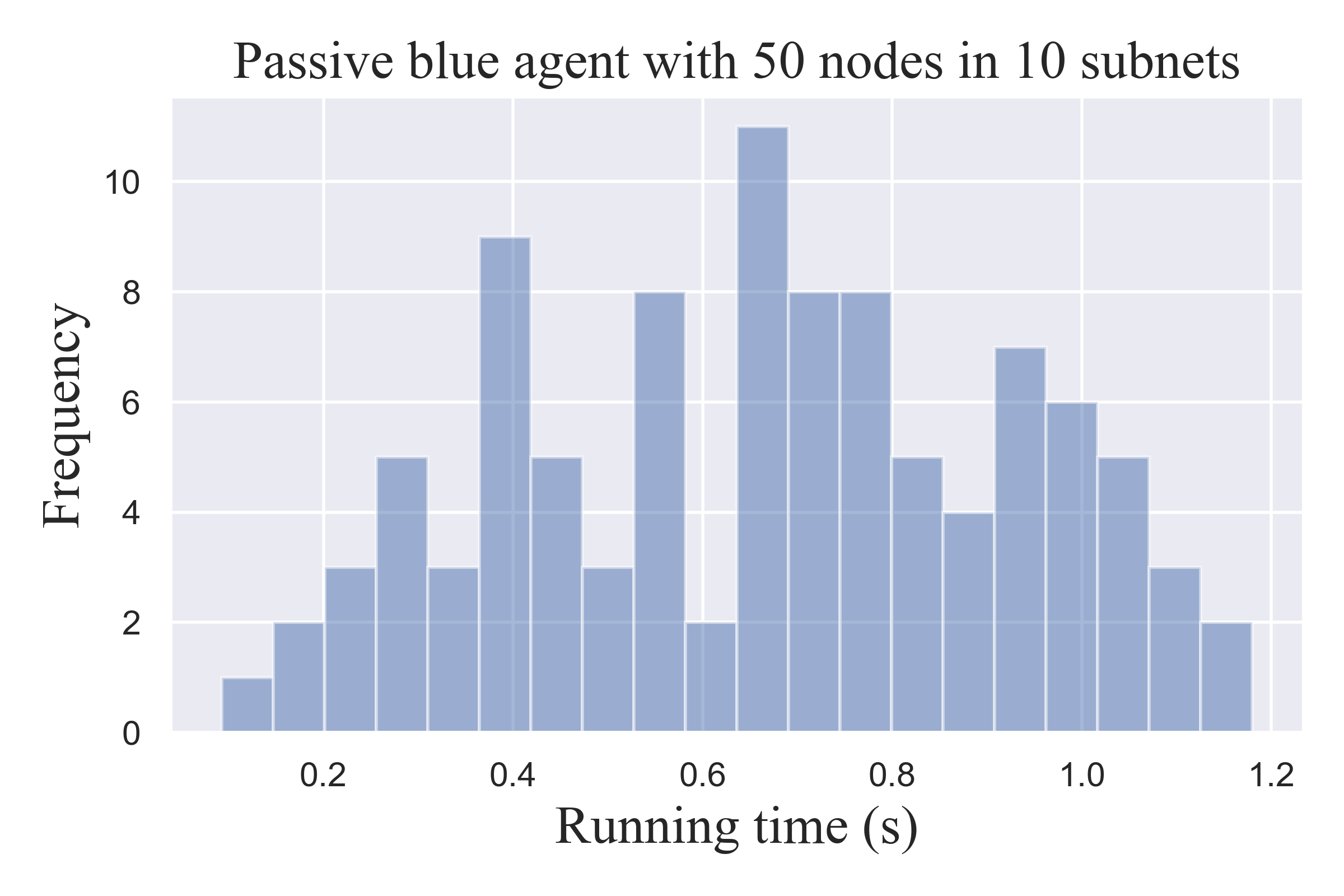} &
    \includegraphics[width=.24\textwidth,height=3.4cm]{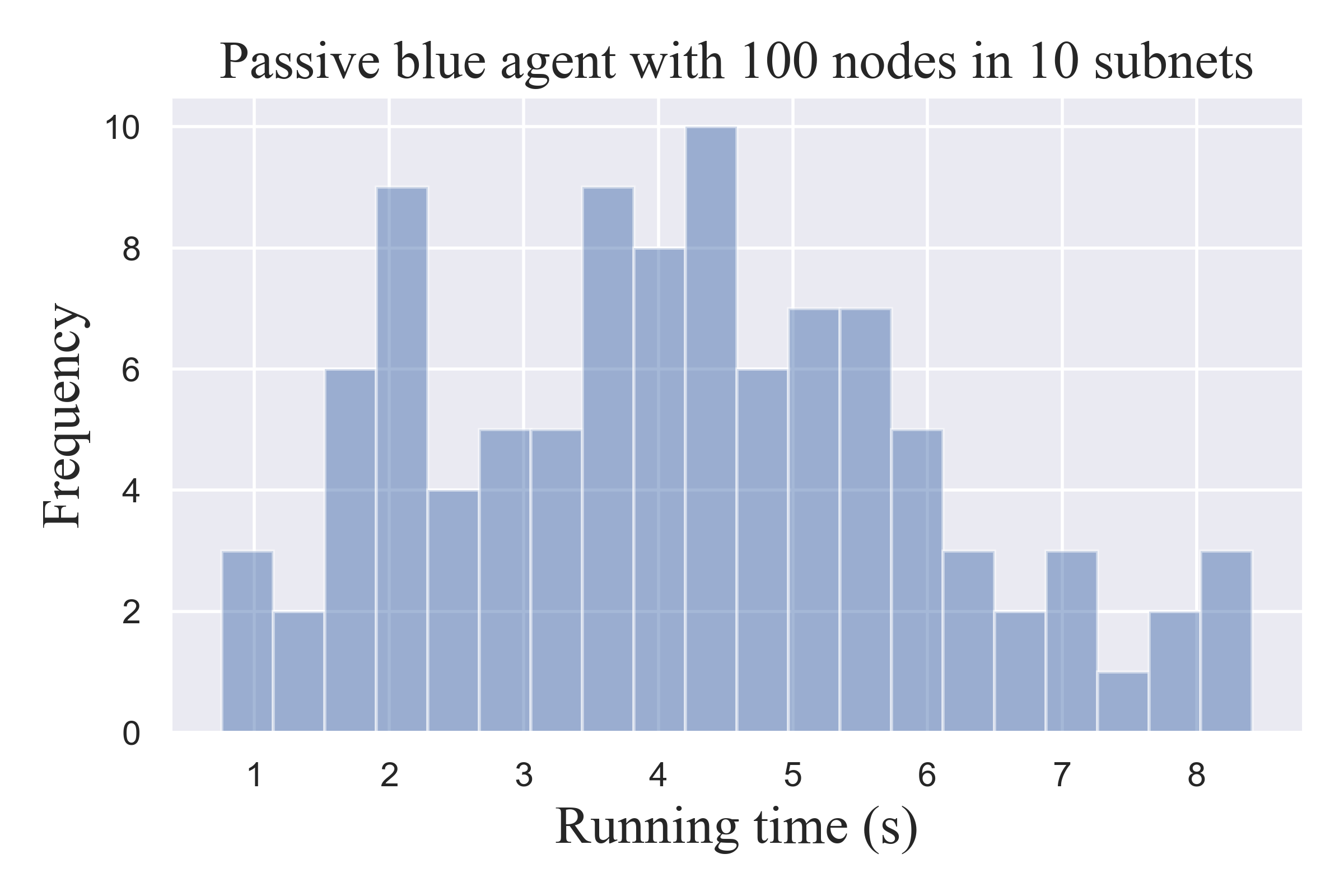} &
    \includegraphics[width=.24\textwidth,height=3.4cm]{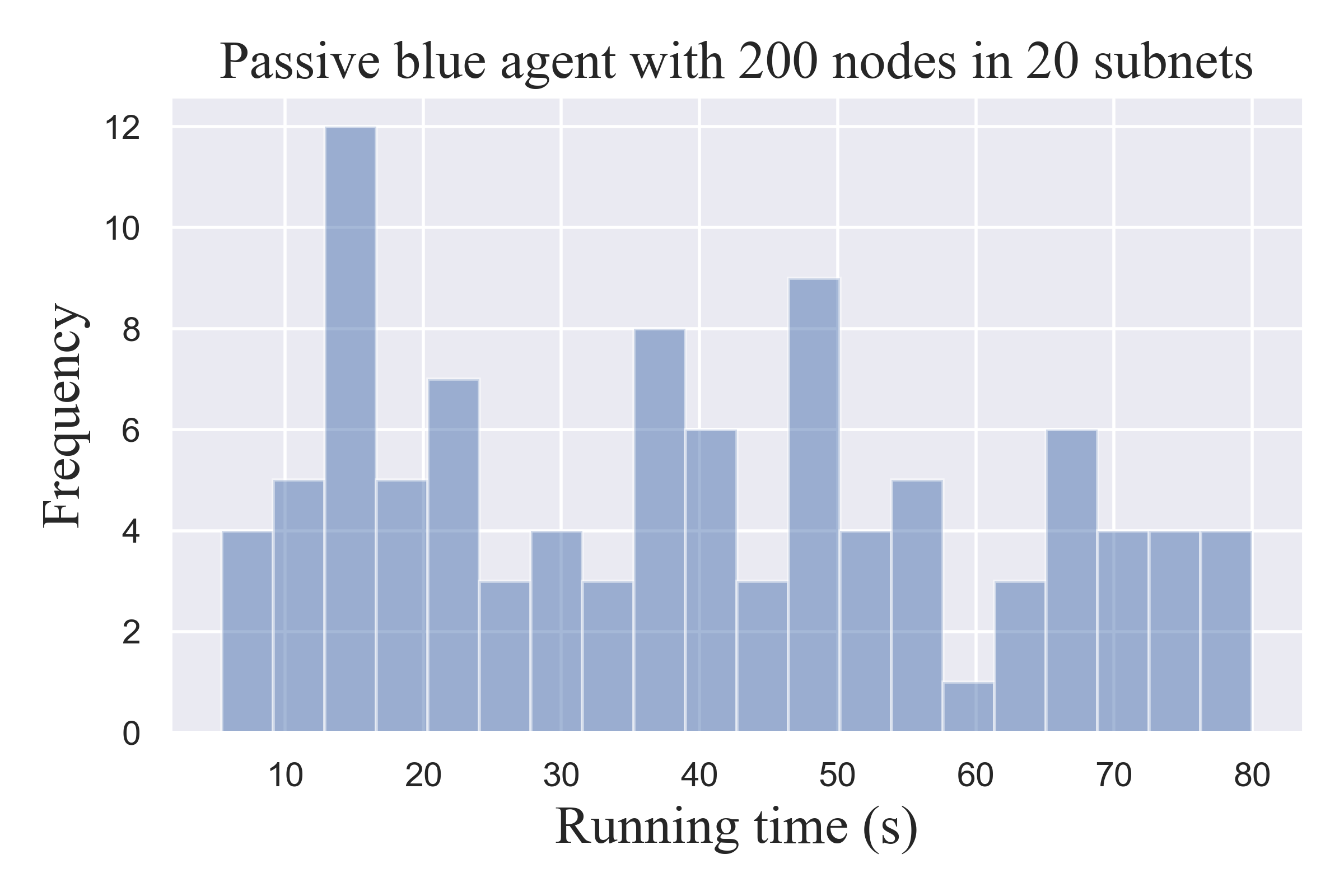}
\end{tabular}
\caption{Frequencies of episode running times for configurations with 25, 50, 100 and 200 nodes respectively. Each histogram describes 100 episodes for each configuration with a passive blue agent.}
\label{fig:episode_plots}
\end{figure*}

Another important aspect of evaluating performance is understanding how long it
would take to learn such a policy. DRL typically requires running a large number
of training episodes. While FARLAND's network emulation layer was designed to
enable the fast network deployment and teardown, it is FARLAND's compatible
simulation layers that make learning to defend a network using DRL practical.
Run time is a consequence of the kind of attack that is simulated, which, in
many cases, involves a computationally expensive searching algorithm (e.g., to
find a server to compromise or file to exfiltrate). Running an episode in
FARLAND can take from a fraction of a second to a few minutes, depending on the
size of the network and other parameters.

Figure~\ref{fig:episode_plots} shows the distribution of running times for 100
episodes on a single \emph(Xeon E5-2660v4) computing core (with a frequency
ranging from 2 to 3.2 GHz). The times correspond to episodes where the red
agent's goal is to find and exfiltrate sensitive content. The blue agent is
passive and does not perform any action, allowing the red agent to achieve its
goal. Despite FARLAND's emulation being considerably faster than emulation
solutions~\cite{noauthor_caldera._2019}, emulated games still take several
orders of magnitude longer than simulation games. For example, games in a 16 CPU
system with the same agent configuration as those in
Figure~\ref{fig:episode_plots} average 10 minutes (with an average setup time of
82 seconds) when emulating networks with 10 nodes. The running time could be
over 2 hours when we emulate networks with 100 nodes or more.

%% file: 12_conclusion.tex
\section{Future work}

FARLAND provides researchers with a tool to design custom network environments
to develop RL cyber defenders. In future work, we plan to address goals
requiring additional features from FARLAND. These goals include the development
of: (i) protection mechanisms against poisoning and evasion attacks; (ii)
assurance evaluation approaches to determine the extent to which a RL-enabled
defender is protected against adversarial manipulation; and, (iii) the
integration of correct (blue) behavior specification. Concretely, a blue agent
would only be practical if network configuration updates preserve packet
traversal invariants (e.g., region segregation, or network-function chains that
preserve partial order). Thus, (iii) relates to the integration of traversal
policy specification into the design of an environment so that blue agents only
perform actions that do not violate packet traversal policies.

To close the gap between simulation and emulation, we plan to bootstrap learning
using smaller networks. Policies can be gradually improved by moving to larger
networks and/or emulated networks. We also plan to optimize the setup and
teardown of networks. 

Finally, we will look into efficient approaches to alternate between simulation
and emulation modes. In particular, our architecture is compatible with
approaches, such as OpenAI's Automatic Domain
Randomization~\cite{openai_solving_2019}. We are exploring the automatic
increase of environment complexity by adjusting probabilistic models when a blue
agent achieves acceptable performance in simpler environments.

\section{Conclusion}

We described FARLAND, a framework to enable research to apply RL to defend
networks. While previous work has demonstrated that RL can be used to learn to
perform complex tasks (such as playing strategy games) with human-level
performance, it has been challenging to translate these achievements to use RL
to perform cybersecurity tasks. 

FARLAND allows researchers to design environments under assumptions that more
closely correspond to real threats. In particular, researchers can develop
attacks where the adversary manipulates observations through realistic actions
in the environment to deceive the defender. In contrast, previous work attacking
RL either assumes that it is possible to directly perturb observations and
rewards~\cite{han_reinforcement_2018}, or studies attacks to RL in a context
that is drastically different from using RL for cybersecurity (e.g.,
\cite{kos_delving_2017,xiao_characterizing_2019}).

Our framework helps to demonstrate that it is feasible to use RL to defend a
network. More importantly, FARLAND is designed to allow cybersecurity
researchers to evaluate their approaches over a range of network conditions that
can vary dramatically, with respect to their complexity and security policies.
FARLAND enables such evaluation and the continuous development of new approaches
as threats evolve.